\documentclass[aps,prb,twocolumn,nofootinbib, superscriptaddress,10pt,floatfix]{revtex4-2}
\usepackage{amsmath,amsthm,amsfonts,amssymb,amscd,bbold,mathrsfs,bm}
\usepackage[T1]{fontenc}
\usepackage{dcolumn}
\usepackage{physics}
\usepackage{graphicx}
\usepackage{epstopdf}
\usepackage{pgf}
\usepackage{tabularx}
\usepackage{subfigure}
\usepackage{float}
\usepackage{hyperref}

\begin{document}
\sloppy

\title{Anisotropic spin-blockade leakage current in a Ge hole double quantum dot}
\author{Zhanning Wang, S. D. Liles, Joe Hillier, A. R. Hamilton, Dimitrie Culcer}
\affiliation{School of Physics, The University of New South Wales, Sydney, New South Wales 2052, Australia}
\date{\today}

\begin{abstract}
Group IV quantum dot hole spin systems, exhibiting strong spin-orbit coupling, provide platforms for various qubit architectures.
The rapid advancement of solid-state technologies has significantly improved qubit quality, including the time scales characterizing electrical operation, relaxation, and dephasing.
At this stage of development, understanding the relations between the underlying spin-orbit coupling and experimental parameters, such as quantum dot geometry and external electric and magnetic fields, has become a priority.
Here we focus on a Ge hole double quantum dot in the Pauli spin blockade regime and present a complete analysis of the leakage current under an out-of-plane magnetic field.
By considering a model of anisotropic in-plane confinement and $k^3$-Rashba spin-orbit coupling, we determine the behaviour of the leakage current as a function of detuning, magnetic field magnitude, interdot distance, and individual dot ellipticities.
We identify regions in which the leakage current can be suppressed by quantum dot geometry designs.
Most importantly, by rotating one of the quantum dots, we observe that the quantum dot shape induces a strongly anisotropic leakage current.
These findings provide guidelines for probing the spin-orbit coupling, enhancing the signal-to-noise ratio, and improving the precision of Pauli spin blockade readout in hole qubit architectures.
\end{abstract}
\maketitle

\section{Introduction}
\label{Section 1 - Introduction}
Qubits based on hole spins in semiconductor quantum dots have received extensive attention in quantum computing research due to their inherent scalability and compatibility with quantum interfaces and devices \cite{Loss1998, Kane1998, Ladd2002, Kalra2014, Salfi2016, Jehl2016, Veldhorst2017, Yoneda2018, Landig2018, Samkharadze2018, Watzinger2018, Scappucci2021, Fang2022, Burkard2023, Yu2023, Ungerer2024, Chien2024, Floor2024}.
In particular, single hole spin qubits in group IV elements like germanium and silicon have demonstrated fast and purely electrical spin manipulations and large electrical modulation of $g$-factors due to strong spin-orbit coupling (SOC) \cite{Ares2013, Liles2018, Mizokuchi2018, Hardy2019, Wei2020, Hendrickx2020, Liles2021, Froning2021_NN, Ke2022, Piot2022}.
In pioneering studies of electron qubits, SOC was found to be weak, necessitating enhancement via micromagnets.
This introduced experimental complications and scalability limitations \cite{Bluhm2011, Pla2012, Takeda2016, Veldhorst2015, Watson2018, Yoneda2021}.
These obstacles are mitigated in hole systems, which possess intrinsically strong SOC due to their $p$-orbital structure in the valence band \cite{Bychkov1984, Rashba1988, Rashba2003, Winkler2003, Culcer2006, Hong2018, James2021, Sina2023}.
Meanwhile, the absence of valley degeneracy avoids complications associated with the enlarged parameter space in electronic qubits \cite{McGuire2007, Chutia2007, Cywinski2010, Culcer2010, Friesen2010, Hao2014, Boross2016, Niquet2018}.

Motivated by these advantages, research on hole systems has led to numerous experimental advancements, such as tunable $g$-factors in nanowire and quantum dot systems \cite{Ares2013_PRL, Pribiag2013, Schroer2011, Voisin2016, Froning2021, Ruoyu2020, Hendrickx2020_NC}, high temperature qubit operation \cite{Shimatani2020, Camenzind2022, Huang2024}, dispersive readout \cite{Lada2018, Crippa2019, Duan2021, Ezzouch2021, Russell2023}, qubit-photon interference \cite{Yu2023, Ungerer2024}, coupling between superconducting microwave resonators and qubits, as well as qubit spin shuttling \cite{Li2018,Xu2020, Nigro2024, Floor2024, DeSmet2024}, and quantum logic circuits \cite{Hendrickx2021}.
Progress in semiconductor quantum dot electron and hole qubits has also stimulated the development of alternative schemes, such as donor-acceptor qubits \cite{Fuhrer2009, Krauth2022, Takashi2018, He2019, Kobayashi2021}, singlet-triplet transition qubits \cite{Levy2002, Petta2005, Maune2012, Jirovec2021, Jirovec2022, Bopp2025}, exchange qubits \cite{Fei2015, Wang2016_2, Geyer2024}, and hybrid qubits \cite{Thorgrimsson2017, Corrigan2023}.

In parallel with advancements in the experimental realization of quantum dot hole devices, theoretical work over the past several decades has supported experiments in optimizing the coherence properties of various qubits and established a solid understanding of these mesoscopic systems \cite{Luttinger1955, Rashba1988, Winkler2000, Khaetskii2002, Winkler2002, Winkler2004, Fischer2010, Bosco20212, Kloeffel2013, Bosco2022_PRL, Lidal2023, Michal2023, Dey2024}.
In search of low noise and rapid qubit operations, efforts have focused on investigating the mechanisms of SOC in quantum confinement structures \cite{Kloeffel2011, Durnev2014, Miserev20172, Marcellina2018, Kloeffel2018}.
This includes both linear-type direct Rashba SOC in nanowire quantum dots and cubic-type Rashba SOC in quasi-two-dimensional quantum dots \cite{Wang2021, Milivojevic2021, Bosco20213, Abhik2023}, paving the way for improved control of electron-dipole spin resonance (EDSR) \cite{Golovach2006, Bulaev2007, Terrazos2021}.
Magnetic field control of electrically driven hole spin qubits is also being investigated, demonstrating improved scalability in planar hole spin qubit structures \cite{Martinez2022, Adelsberger20222, Bosco2022, Malkoc2022, Wang2024}.

Beyond electrical and magnetic control of qubits, strain is increasingly recognized as an important factor \cite{Liles2021}.
Axial and shear strain are believed to enhance the EDSR rate and influence the dephasing time in non-trivial ways \cite{Venitucci2020, Carlos2022, Mena2023}, while non-uniform strain arising from material stacking and gate electrodes have recently been reported as important factors in SOC modulation \cite{Shalak2023, Sarkar2025}.
Beyond optimizing qubit scalability through external fields and strain engineering, quantum dot geometries are being examined in detail to enhance SOC in elliptical quantum dots due to orbital effects.
This approach offers further potential for achieving faster EDSR rates and improved dephasing times \cite{Niquet2019, Bosco2021, Adelsberger2022, Shalak2023, Wang2024, Chien2024_2}.

From coherent qubit control to signal readout, precise control of $g$-factors and determination of SOC play important roles in both experimental and theoretical studies.
The Pauli spin blockade (PSB) phenomenon offer an ideal pathway to address these challenges by allowing dispersive signal read-out, thereby simplifying the device structure \cite{Danon2013, Wang2016, Qvist2022, Lundberg2024}.
The leakage current has been widely used for spin readout in transport experiments \cite{Ono2002, Ono20024, Mizokuchi2018, Hendrickx2020_NC}, as a method based on PSB.
Leakage mechanisms play a similar role in PSB readout and have already proven effective in extracting the spin relaxation time of a single hole spin qubit \cite{Lai2011, Li2015, Pla2012}.
In recent experiments, the spin-to-charge conversion enabled by PSB has been adapted into dispersive readout schemes, further improving signal quality \cite{Betz2015, Samkharadze2018}.
A standard PSB setup involves a double quantum dot configuration with two carriers (electrons or holes).
If both carriers have the same spin, they will form a triplet state, and tunneling is blocked by the Pauli exclusion principle; if they have different spins, they form a singlet state, allowing a finite current.
In hole systems, strong SOC enables spin flips that lift the triplet blockade, introducing a leakage current and reducing PSB readout fidelity \cite{Bohuslavskyi2016, Zarassi2017, Jirovec2021, Jirovec2022, Liles2024}.
Earlier theoretical studies have demonstrated a strong dependence of the leakage current on the magnetic field orientation, predicting a vanishing leakage current when the magnetic field is parallel to the SOC vectors \cite{Qvist2022_2}, and the polarization states of the lead connecting the quantum dots have been proven to be an important tool for characterizing the $g$-factors of the quantum dots \cite{Mutter2021}.
{Furthermore, recent experimental advances have demonstrated that an in-plane magnetic field enables strong control over the coupling between two quantum dots about $50\%$, opening a new pathway to tune charge tunneling even after device fabrication is completed \cite{Bopp2025}.}
A key requirement linking experimental results and theoretical approaches is understanding the dependence of the leakage current on the quantum dot geometry.
Achieving this understanding is the central aim of this work.
\begin{figure}[t!]
    \centering
    \includegraphics[width=1\columnwidth]{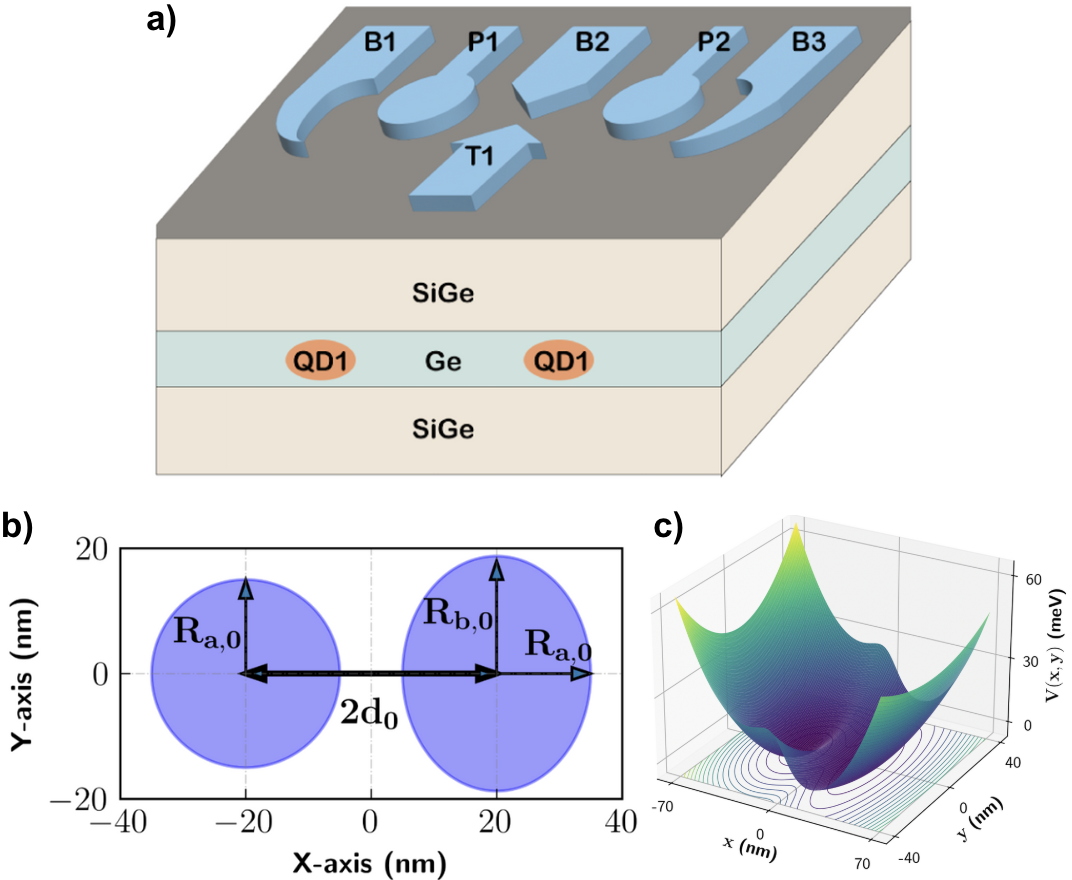}
    \caption{
    A schematic view of a Ge double quantum dot system.
    a) The full double quantum device.
    The substrate includes a fully strain-relaxed SiGe layer at the bottom.
    The middle of the heterostructure consists of an epitaxially grown layer of strained germanium, hosting the hole qubit, and another layer of relaxed SiGe atop the Ge layer.
    Gates B1, P1, B2, P2, and B3 are used to confine two quantum dots, while gate B2 and T1 can control the inter-dot tunneling and detuning parameter $\delta$.
    b) A planar view of the double quantum dot system in the $xy$-plane.
    The two quantum dots are depicted as shaded solid circles.
    The left quantum dot, located at $x=-d_0$, is a circular dot with radius $R_{a,0}$, while the right quantum dot, located at $x=d_0$, is elliptical, with semi-minor axis of $R_{b,x}$ and semi-major axis of $R_{b,y}$.
    The two quantum dots are separated by a distance of $2d_0$, referred to as the interdot distance in the following text.
    c) {A three-dimensional view of the double quantum dot potential function described in Eq.~\eqref{Eq: DOD Potential}.
    The ellipticity of the confinement potential, corresponding to b), is reflected in the contour projection on the $V(x,y)=0$ surface.}
    }
    \label{Fig: Ge double quantum dot}
\end{figure}

In this article, we consider a germanium double quantum dot system in the vicinity of the PSB regime in an out-of-plane magnetic field.
We model the system using the quasi-two-dimensional Luttinger-Kohn Hamiltonian within the effective mass approximation.
{Based on the well-studied case of spin blockade in two-dimensional electron gas systems \cite{Fransson2006, Amo2007,Brauns2016, Seedhouse2021}, we have adopted an explicit model for the cubic-Rashba spin–orbit coupling derived from the four-band Luttinger–Kohn Hamiltonian, as well as the strain-induced linear-Rashba spin–orbit coupling.}
Considering the complexities of spin blockade in spin-3/2 systems, the first step in achieving a systematic understanding of the leakage current is to elucidate the underlying charge and spin dynamics in a hole system subject to a simple, homogeneous, strain configuration.
Our work focuses on precisely such a case -- assuming uniform axial and shear strain based on experimental data.
Whereas inhomogeneous strain is known to play an important role in qubit dynamics \cite{Carlos2022} and must eventually be taken into account, which can in principle be done using the approach of Ref.~\cite{Sarkar2025}.
Nevertheless, we stress that the PSB leakage current is essentially the result of a two-qubit calculation.
The inclusion of a sufficient number of excited states to achieve convergence would result in a Hilbert space several orders of magnitude larger than that of Ref.~\cite{Sarkar2025}, which is challenging at present.
For this reason the study of two spin-3/2 hole qubits in the presence of inhomogeneous strain is necessarily a longer-term undertaking.

In this work we focus on two quantum dot geometries.
In the first case, both quantum dots are circular with the equal radii ($R_{a}$=$R_{b}$).
In the second case, the left quantum dot is circular ($R_{a,x}$=$R_{a,y}$=$R_{a,0}$), while the right quantum dot is elliptical, with a semi-minor axis of $R_{a,0}$ and a semi-major axis ($R_{b,y}=R_{a,0}$).
Such anisotropic confinement can enhance the SOC \cite{Qvist2022, Dey2024}.
Furthermore, the elliptical quantum dot is rotated and swept through an angle of $\pi$.
Our main findings are as follows:
(1) For the case of one circular quantum dot and one elliptical quantum dot, we find a sharp peak in the leakage current as a function of the detuning (the relative energy between the two quantum dots).
Analytical expressions for both the leakage current and the peak location are given below.
(2) The leakage current is a non-monotonic function of the magnetic field, exhibiting a smooth maximum.
(3) In the special case where two quantum dots are identically circular, there will be no leakage current, and the PSB cannot be lifted via electric or magnetic control.
By examining the leakage current as a function of the interdot distance and the aspect ratio ($e=\omega_y/\omega_x$) of the elliptical quantum dot, we attribute the non-monotonic behavior from the perspective of the interplay between SOC and the wave function overlap of the two dots.
(4) Our most important result is the strong anisotropy in the leakage current as the axis of the elliptical quantum dot is rotated.
We find that when the semi-major axis is parallel to the $x$-axis, the leakage current is minimized.
These findings may suggest experiments aimed at understanding the site-dependent $g$-factors, SOC, PSB, and at improving coherent control of hole qubits.

The outline of this paper is as follows.
In Sec.~\ref{Section 2 - Model and Methodology}, we introduce the model of the germanium hole double quantum dot system. Then, within the framework of Hund-Mulliken molecular orbital theory, we present the wave functions for both the circular and elliptical quantum dots, which are key to diagonalizing the double quantum dot Hamiltonian.
This is followed by the derivation of the leakage current in the PSB regime using the Lindblad quantum kinetic equation for the $(1,1)-(0,2)$ qubit space.
In Sec.~\ref{Section 3 - Results and Discussions}, we present the energy levels in various parameter regimes, paving the way to analyze the leakage current.
We then evaluate the leakage current as a function of detuning and magnetic field, pointed out the role of spin-orbit coupling matrix elements.
Finally, we discuss the strong anisotropy of the leakage current as a function of relative quantum dot orientations.
The last section Sec.~\ref{Section 4 - Conclusion and Outlook} concludes this work with suggestions for possible outlooks.

\section{Model and Methodology}
\label{Section 2 - Model and Methodology}
We consider a double quantum dot system in a germanium hole heterostructure.
Germanium offers several advantages in quantum dot fabrication, including the possibility of isotopic purification \cite{Itoh1993}, potential suitability for strain engineering, and higher mobilities \cite{Sammak2019, Lodari2019, Lawrie20202, Stehouwer2023, Corley-Wiciak2023}.
However, the method developed in this work can also be adapted to a silicon metal–oxide–semiconductor platform.
A schematic model of the double quantum dot configuration in germanium is depicted in Fig.~\ref{Fig: Ge double quantum dot}a).
The gate electric field $F$, is fixed at 10\,MV/m, and the quantum well width $L$ is set to 10\,nm.
The quantum well width $L$ determines the heavy-hole–light-hole splitting, and thus the Rashba SOC coefficients \cite{Marcellina2017_2, Abhik2023}.
A zoomed view of the in-plane structure is shown in Fig.~\ref{Fig: Ge double quantum dot}b).
The left quantum dot couples to a source with tunneling rate of $\Gamma_{\text{in}}$, while the right quantum dot couples to a drain with a tunneling rate of $\Gamma_{\text{out}}$, which determines the scale of the leakage current \cite{Danon2009, Mutter2021}.
In this work, we  use $\Gamma$ as the smaller value between the $\Gamma_{\text{in}}$ and $\Gamma_{\text{out}}$, and $\Gamma$ is set to be $1$\,GHz.
The system is assumed to be calibrated around the (1,1)$\to$(0,2) charge transition regime, where $(n,m)$ indicates an occupation of $n$ holes in the left quantum dot and $m$ holes in the right quantum dot.

Using the parameters given above, the heavy-hole–light-hole (HH-LH) energy splitting is approximately 50 meV, while the energy gap between the ground and excited states is around 7 meV.
Our calculations focus on the case in which the detuning is smaller than the orbital level splitting for each quantum dot.
Therefore, higher orbital excited states are not included in this study due to the large energy gap between excited states \cite{Burkard1999, Hanson2007, Qiuzi2010}.
Additionally, thermally induced tunneling processes are disregarded.
Consequently, only five states are considered in the singlet-triplet transition process: three (1,1)-triplet states, denoted as $\ket{T_+},\,\ket{T_0},\,\ket{T_-}$, one (1,1)-singlet state, denoted as $\ket{S_{(1,1)}}$, and one (0,2)-singlet state, denoted as $\ket{S_{(0,2)}}$.
Spin-preserving tunneling to the $\ket{S_{(0,2)}}$ state is allowed when the system is initially in the $\ket{S_{(1,1)}}$ state (i.e. charge tunneling).
With the help of SOC, spin-flip tunneling from the (1,1)-triplet states to the $\ket{S_{(0,2)}}$ state becomes possible, thereby lifting the PSB \cite{Hanson2007, Zwanenburg2013}.
The central aim of this work is to understand the factors that may affect the PSB by investigating the leakage current $I$, including the detuning parameter $\delta$, the out-of-plane magnetic field $B_z$, and the quantum dot geometries.

This section is organized as follows.
In Sec.~\ref{Section 2 - Subsection 1 - Model Hamiltonian}, we introduce the model Hamiltonian of the double quantum dot system.
Sec.~\ref{Section 2 - Subsection 2 - Hund-Muliken Approach} introduces the construction of two-hole wave functions based on the Hund-Mulliken molecular orbital approach, and the details of the wave functions for a rotated elliptical quantum dot are also discussed.
In Sec.~\ref{Section 2 - Subsection 3 - Leakage current}, the expressions for the leakage current is derived using the Lindblad quantum kinetic equation.
\subsection{Model Hamiltonian}
\label{Section 2 - Subsection 1 - Model Hamiltonian}
The (1,1)$\to$(0,2) charge transition in this work is described by the total Hamiltonian $H=H_0 + H_{\text{SO}} + H_{\text{Z}} + V_{\text{C}}$.
The quantum dot Hamiltonian in the two-hole basis can be written as $H_0=H_{1,\text{SQD}} + H_{2,\text{SQD}}$, where each term is implicitly extended to the full two-hole Hilbert space by tensoring with the identity operator on the complementary subspace \cite{Burkard1999}.
Using $i$ to index the number of holes, the single-hole Hamiltonian for each quantum dot is denoted as:
\begin{equation}\label{EQ: Single Quantum Dot Hamiltonian}
    H_{i,\text{SQD}}=\frac{\pqty{\bm{p}_i+q\bm{A}(\bm{r}_i)}^2}{2m_{\text{HP}}} + V(\bm{r}_i) \,.
\end{equation}
Here, $m_{\text{HP}}$ denotes the in-plane effective mass of the heavy hole (HH), $\bm{A} = \pqty{-B_z y, B_z x, 0}$ is the vector potential, and $q$ is the charge of a hole.
The in-plane confinement potential is constructed based on the assumption that the confinement potentials are parabolic near the center of each quantum dot.
To smoothly connect the parabolic confinements of the left and right quantum dots, we use a $\tanh$ function for $V(\bm{x})$:
\begin{equation}
\begin{aligned}\label{Eq: DOD Potential}
    & \frac{1}{4}m_{\text{HP}}\omega_{a,0}^2\pqty{(x_i-d_0)^2-(x_i+d_0)^2} \tanh{\frac{x}{\lambda_0}}\\
    & +\frac{1}{4}m_{\text{HP}}\omega_{a,0}^2\pqty{(x_i-d_0)^2+(x_i+d_0)^2}\\
    & -\pqty{\frac{1}{4}m_{\text{HP}}\omega_{a,0}^2y_i^2 - \frac{1}{4}m_{\text{HP}}\omega_{b,y}^2y^2} \tanh{\frac{x_i}{\lambda_0}} +\\
    & +\frac{1}{4}m_{\text{HP}}\omega_{a,0}^2y_i^2+\frac{1}{4}m_{\text{HP}}\omega_{b,y}^2y_i^2 \,.
\end{aligned}
\end{equation} 
The confinement frequencies are $\omega_{a,0} = \pqty{m_{\text{HP}} R_{a,0}^2 / \hbar}^{-1}$ and $\omega_{b,y} = e \omega_{b,x}$.
The free parameter $\lambda_0$ is chosen to be 10~nm, where a larger $\lambda_0$ indicates faster convergence of the confinement to a parabolic potential near the barrier region, i.e., as $x \to 0$.
Previous studies have focused on quartic confinement potentials (involving $x^4$ and $y^4$ terms), which leads to stronger confinement energies at large interdot distances.
Furthermore, the barrier height at the origin of the model used here is $m_{\text{HP}}\omega_0^2d_0^2/2$, which is four times higher than a quartic potential \cite{Burkard1999, Qiuzi2010}.
The gate electric field-induced Rashba SOC term, $H_{\text{SO}}$, has the form:
\begin{equation}\label{EQ: HSO}
    H_{\text{SO}}=\Omega_{1,\,\text{SO}} + \Omega_{2,\,\text{SO}} \,.
\end{equation}
$\Omega_{1,\text{SO}}$ and $\Omega_{2,\text{SO}}$ are derived, following a similar approach introduced in Ref.~\cite{Marcellina2018}, from the four-band Luttinger-Kohn Hamiltonians using the Peierls substitution: $\bm{k}_i \to \bm{k}_i + e\bm{A}(\bm{r}_i)$.
Now, focusing on the single-hole basis, in the quasi-two-dimensional limit, the Rashba SOC $\Omega_{i,\text{SO}}=$ takes the form:
\begin{equation}
\begin{aligned}
   & i \alpha_{R2}\left(k_{i,+}^3 \sigma_{i,-}-k_{i,-}^3 \sigma_{i,+}\right) +\\
    &i \alpha_{R3}\left(\left\{k_{i,+}, k_{i,-}^2\right\} \sigma_{i,-}-\left\{k_{i,+}^2, k_{i,-}\right\} \sigma_{i,+}\right) +\\
    & i \alpha_{R1} (k_{i,x} \sigma_{i,y} - k_{i,y} \sigma_{i,x})\,.
\end{aligned}
\end{equation}
The wave vectors are defined as $k_\pm=k_x\pm ik_y$, where $k_x$ and $k_y$ carry the orbit magnetic field.
$\sigma_\pm=\sigma_x \pm i \sigma_y$, where $\sigma_x$ and $\sigma_y$ are the standard Pauli X and Y matrices.
In this work, we have fixed the gate electric field at 10~MV/m and the quantum well width at 10~nm.
The coefficients $\alpha_{R2}$ and $\alpha_{R3}$ represent the cubic Rashba SOC terms \cite{Moriya2014,Marcellina2017_2}.

These coefficients were initially studied for single-hole EDSR operations, as they can be enhanced by both the gate electric field and the quantum well width \cite{Golovach2006, Bulaev2007}.
Their magnitude strongly depend on the strain components (estimated following Refs.~\cite{Terrazos2021, Abhik2023}).
The $\alpha_{R1}$ Rashba SOC term exhibits a linear dependence on $k$, and is adapted from Refs.~\cite{Carlos2022, Yang2022} for systems with constant shear strain.
This term incorporates the effects of atomistic potential-induced shear strain and non-uniform strain distributions, which have been shown to play a critical role in enhancing qubit coherence times through strain engineering and gate electrode design.
However, a comprehensive exploration of all possible strain effects is beyond the scope of this work \cite{Shalak2023, Mauro2024, Mauro2024_2, Sarkar2025}.

Our model does not include bulk inversion asymmetry (bulk Dresselahus terms) or interface roughness induced asymmetry (surface Dresselhaus terms) \cite{Durnev2014}.
In group III-V hole devices, both theoretical and experimental studies suggest that the Dresselhaus term contributes significantly to the leakage current \cite{Wang2016, Hung2017}.
In group IV hole-based devices bulk Dresselhaus terms are absent, while systematically incorporating surface Dresselhaus terms requires device-specific modeling.
We thus defer a systematic study of them to future works.
The direction of the SOC is comprehensively discussed in Ref.~\cite{Qvist2022_2}, which demonstrates that when the magnetic field is aligned with the spin-orbit direction, the PSB is restored, leading to vanishing leakage current.
Furthermore, recent experiments in Ref.~\cite{Liles2024} also reported a marginal out-of-plane component of the SOC matrix element; combined with the findings of Ref.~\cite{Qvist2022_2}, the consideration of an out-of-plane component in the SOC is excluded.

The Zeeman Hamiltonian, considering only the out-of-plane magnetic field, reads:
\begin{equation}
H_{\text{Z}} = g_1 \mu_B B_z \sigma_{1,z} + g_2 \mu_B B_Z \cdot \sigma_{2,z} \,,
\end{equation}
where $g_1$and $g_2$ are the effective $g$-factors at the locations of hole 1 and hole 2, respectively.
For example, if hole 1 is in the left (right) quantum dot, $g_1$ will correspond to $g_L$ ($g_R$).
In general, the $g_L$ and $g_R$ are tensors that depend on left or right quantum dot properties, such as local gate geometry, interface roughness, and strain profiles.
These microscopic mechanisms can lead to HH-LH mixings and 
A full modeling of the individual $g$-tensor effects requires a multiband analysis, which is beyond the scope of our current leakage current calculations.
Therefore, we adopt a phenomenological description of the $g$-factors, using experimental fittings reported in Ref.~\cite{Jirovec2021}.
The two-body Coulomb interaction is denoted as $V_{\text{C}}$:
\begin{equation}
V_{\text{C}}=\frac{1}{4 \pi \epsilon_0 \epsilon_r} \frac{e^2}{\norm{\bm{x}_1-\bm{x}_2}} \,,
\end{equation}
where $\epsilon_0$ is the vacuum permittivity and $\epsilon_r=15.8$ is the relative permittivity in germanium.
With the full Hamiltonian now established for calculating the leakage currents, the next step is to find a set of basis wave function to determine the spectrum of the Hamiltonian $H$.

\subsection{Hund-Muliken Approach}
\label{Section 2 - Subsection 2 - Hund-Muliken Approach}
In this double quantum dot structure, two holes become delocalized from the individual quantum dot orbitals due to the finite height of the potential barrier at origin, and the SOC Hamiltonian $H_{\text{SO}}$.
Consequently, the two-hole wave functions can be obtained using Hund-Mulliken molecular orbital theory, which is constructed from the single-hole wave functions \cite{Mulliken1927, Mulliken1967, Burkard1999, Culcer2009, Qiuzi2010}.
The first step is to determine the single-hole quantum dot wave functions around $x=\mp d_0$ according to Eq.~\eqref{EQ: Single Quantum Dot Hamiltonian}.
Starting with the left circular quantum dot centered at $x=-d_0$, locally, the potential degenerates into a standard parabolic confinement.
In this case, the wave function for a hole in the left quantum dot, labelled by subscript $L$ will read:
\begin{small}
\begin{equation}
    \phi_L(\bm{x})=(R_L\sqrt{\pi})^{-1} \exp \left(-\frac{\left(x+d_0\right)^2+y^2}{2 R_L^2}+i \frac{d_0y}{2 R_B^2}\right) \,,
\end{equation}
\end{small}
where $R_L=\pqty{m_{\text{HP}} \Omega_L / \hbar}^{-1/2}$ is the effective dot radius modified by the magnetic field $B_z$, and $\Omega_L=\sqrt{\omega_{L,0}^2 + \omega_c^2 / 4}$.
We have performed a gauge transformation to the quantum dot Hamiltonian, which is reflected in the phase factor $\exp \left(\mathrm{i}\,d_0y/2 R_B^2\right)$, where $R_B=\pqty{m_{\text{HP}} \omega_c / \hbar}^{-1/2}$.
To understand how the wave function changes with the magnetic field, we observe that $R_L$ decreases as $B_z$ increases, while $R_B$ shows the opposite trend.

The wave function for the right quantum dot, labelled by subscript $R$, is more complex due to its ellipticity.
We adopt the method used in Ref.~\cite{Avetisyan2012} to obtain the wave function for an elliptical quantum dot:
\begin{small}
\begin{equation}
\begin{aligned}
  \phi_R(\bm{x})=& (\pi R_{R,x}R_{R,y})^{-1/2} \exp(-i\frac{y d_0}{2 R_B^2}) \times\\
  &\exp \left[-\frac{\left(x-d_0\right)^2}{2 R_{R,x}^2}-\frac{y^2}{2 R_{R,y}^2}-i \frac{y\left(x-d_0\right)}{R_{R,xy}^2}\right] \,.
\end{aligned}   
\end{equation}
\end{small}
We notice that the factors $R_{R,x}$ and $R_{R,y}$ represent the effective quantum dot sizes, depending on $\omega_{R,x}$ and $\omega_{R,y}$, respectively.
The full expressions for $R_{R,x}$ and $R_{R,y}$ can be found in the supplementary materials.
To compare the influence of quantum dot anisotropy on the leakage current, the parameter $\omega_{R,x}$ is fixed to be the same as that of the circular quantum dot on the left, and we introduce the aspect ratio $e^2=\omega_{b,y} / \omega_{a,x}$ to control the aspect ratio of the right quantum dot.
A smaller $e$ indicates a weaker confinement frequency or, equivalently, a larger semi-axis radius along the $y$-axis.
After obtaining the single quantum dot wave functions, the molecular orbital states can be constructed as $\ket{L}=(\ket{\phi_L}-g\ket{\phi_R})/\sqrt{1-Sg+g^2}$, and $\ket{R}=(\ket{\phi_R}-g\ket{\phi_L})/\sqrt{1-Sg+g^2}$, where $S=\braket{\phi_L}{\phi_R}$ describes the overlap of the original localized orbital states, and $g(1-\sqrt{1-S^2})/S$.
As the magnetic field increases, the quantum dot size shrinks, leading to a smaller overlap $S$.
The relevant Hilbert space for the charge transition (1,1) $\to$ (0,2) can be spanned by: $\ket{T_+}=\ket{\uparrow,\uparrow}(\ket{L,R}-\ket{R,L})$, $\ket{T_-}/\sqrt{2}=\ket{\downarrow,\,\downarrow}(\ket{L,R}-\ket{R,L})/\sqrt{2}$, $\ket{T_0}=\pqty{\ket{\uparrow\downarrow}+\ket{\downarrow\uparrow}}(\ket{L,R}-\ket{R,L})/2$, $\ket{S_{(1,1)}}=\pqty{\ket{\uparrow\downarrow}-\ket{\downarrow\uparrow}}(\ket{L,R}+\ket{R,L})/2$, and $\ket{S_{(0,2)}}=\pqty{\ket{\uparrow\downarrow}-\ket{\downarrow\uparrow}}\ket{R,R}/\sqrt{2}$ \cite{Culcer2009, Culcer2012}.
In this basis, the full Hamiltonian $H$ in matrix form is given by:
\begin{equation}
    \label{EQ: Total Hamiltonian Matrix Form}
    \mqty[E_0+\varepsilon_{\text{Z}} & t_{\text{SO},1} & 0 & t_{\text{SO},2} & t_{\text{SO},3} \\
    t^*_{\text{SO},1} & E_0 & t_{\text{SO},1} & 0 & 0 \\
    0 & t^*_{\text{SO},1} & E_0-\varepsilon_{\text{Z}} & -t_{\text{SO},2} & -t_{\text{SO},3} \\
    t^*_{\text{SO},2} & 0 & -t^*_{\text{SO},2} & E_0+\Delta u & t_{\text{C}} \\
    t^*_{\text{SO},3} & 0 & -t^*_{\text{SO},3} & t^*_{\text{C}} & E_1] \,.
\end{equation}
The $E_0=\varepsilon_0+u_-$ term includes two parts, the first part is the on-site energy:
\begin{equation}
    \varepsilon_0 = \mel**{L}{H_0}{L}+\mel**{R}{H_0}{R} \,,
\end{equation}
which is identical for all the (1,1)-states; the second part is the matrix elements of the Coloumb interaction:
\begin{equation}
\begin{aligned}
    u_- =& \mel**{L_1R_2}{V_{\text{C}}}{L_1R_2}-\mel**{L_1R_2}{V_{\text{C}}}{R_1L_2} \\
    &- \mel**{R_1L_2}{V_{\text{C}}}{L_1R_2} + \mel**{R_1L_2}{V_{\text{C}}}{R_1L_2} \,,
\end{aligned}
\end{equation}
which is the identical for the (1,1)-triplet states.
For the (1,1)-singlet states, the matrix element for the Coulomb interaction read:
\begin{equation}
\begin{aligned}
    u_- =& \mel**{L_1R_2}{V_{\text{C}}}{L_1R_2}+\mel**{L_1R_2}{V_{\text{C}}}{R_1L_2} \\
    &+ \mel**{R_1L_2}{V_{\text{C}}}{L_1R_2} + \mel**{R_1L_2}{V_{\text{C}}}{R_1L_2} \,.
\end{aligned}
\end{equation}
However, to write the whole matrix in a compact form, we introduce $\Delta u=u_+-u_-$, which is
\begin{equation}
    \Delta u=2\mel**{L_1R_2}{V_{\text{C}}}{R_1L_2}+2\mel**{R_1L_2}{V_{\text{C}}}{L_1R_2}\,,
\end{equation}
so that the matrix element $\mel**{S_{(1,1)}}{H_0+V_{\text{C}}}{S_{(1,1)}}$ is still $\varepsilon_0+u_+$.
In the (0,2)-subspace, the matrix element $E_1=\mel**{S_{(0,2)}}{H_0+V_{\text{C}}}{S_{(0,2)}}$, describs two holes in the same quantum dot, which can be very large due to the Coulomb repulsion for a small quantum dot like in silicon \cite{Qiuzi2010}.
The energy difference between (1,1)-states and (0,2)-state can be controlled electrically by applying a bias voltage. Hence, in the following discussions, the detuning paramter $\delta$ is used to represent this energy difference, $E_0$ and $E_1$ will be incoperated into $\delta$.
The terms $\pm \varepsilon_{\text{Z}}=\mel**{T_\pm}{H_{\text{Z}}}{T_\pm}$ represent the Zeeman matrix elements, and $t_{\text{C}}=\mel**{S_{(1,1)}}{H_0 + V_{\text{C}}}{S_{(0,2)}}$ is the charge tunneling matrix element.
The spin-orbit matrix elements $t_{\text{SO},i},(i \in \Bqty{1,\,2,\,3})$ are obtained by projecting $H_{\text{SO}}$ onto different states, and can be expressed as:
\begin{align}
    t_{\text{SO},1}=&\frac{1}{\sqrt{2}} \pqty{\mel**{L}{H_{\text{SO}}}{L} + \mel**{R}{H_{\text{SO}}}{R}}\,, \\
    t_{\text{SO},2}=&\frac{1}{\sqrt{2}} \pqty{-\mel**{L}{H_{\text{SO}}}{L} + \mel**{R}{H_{\text{SO}}}{R}}\,, \\
    t_{\text{SO},3}=&-\mel**{L}{H_{\text{SO}}}{R} \,.
\end{align}
In the following sections, a rotated elliptical quantum dot on the right is also considered, with the quantum dot rotated by an angle $\theta$ counterclockwisely.
This is achieved by applying the transformations $x \to x\cos\theta-y\sin\theta$ and $y \to y\cos\theta + x\sin\theta$ to both the wave functions and the in-plane confinement potential.

\subsection{Lindblad equation}
\label{Section 2 - Subsection 3 - Leakage current}
The lifting of the PSB can be probed by measuring the leakage current between the double quantum dot and drain, which can be modeled as an open system problem in a lead-dot-lead configuration.
To describe dynamics of the carrier, we use the Lindblad quantum kinetic equation:
\begin{equation}
    \frac{\partial \rho}{\partial t}=-\frac{i}{\hbar}[H,\,\rho]+\frac{\Gamma}{4} \sum_i\left(\hat{L}_i \rho \hat{L}_i^{\dagger}-\frac{1}{2}\Bqty{\hat{L}_i^{\dagger} \hat{L}_i, \rho}\right) \,.
\end{equation}
Here $H$ is the matrix from Eq.~\eqref{EQ: Total Hamiltonian Matrix Form}, $\rho$ the density matrix of the (1,1)-(0,2) charge transition Hilbert space.
The Lindblad operators $\hat{L}_i$ taking the form $\ketbra{i}{S{(0,2)}}$, connect various possible states to the $\ket{S_{(0,2)}}$ state, which allows a hole to flow out of the system via a lead.
$\Gamma$ is assumed to be a constant describing the decay rate to the $\ket{S_{(0,2)}}$ state and the refilling of the $\ket{S_{(1,1)}}$ state.
{The intrinsic spin relaxation is neglected for two resonances.
Spin relaxation channels arising from hyperfine interactions are strongly suppressed in germanium due to isotopic purification of ${}^{73}$Ge, and further reduced by the $p$-orbital character of the valence band.
Another important relaxation channel involves spin–phonon coupling mediated by SOC mechanism.
Unlike group III–V semiconductors, where piezoelectric is also present, shows power law dependence $1/T_1 \sim B^5$ or $B^7$ \cite{Golovach2004, Bulaev2005}.
Group IV hole qubits exhibit significantly longer relaxation times, reaching up to $\sim 30$ ms, which is negligible compared to the tunneling rate \cite{Danon2009, Mutter2021, Lawrie20202, Qvist2022_2}.}

The leakage current for measuring the (1,1)$\to$(0,2) charge transition is given by $I=e \Gamma \mel**{S_{(0,2)}}{\rho}{S_{(0,2)}}$, reducing the problem to evaluating the matrix elements of the density matrix $\rho_{mn}$.
In steady state, where $\partial \rho / \partial t=0$, we firstly vectorize the density matrix $\rho$ into a single column vector.
Now, we get a linear system $A \bm{x} = \bm{b}$, where $A$ comes from the Lindblad operators and and normalization condition $\tr\rho=1$, and $\bm{b}$ will only contain an entry of 1 and the rest are 0.
Performing a formal inversion, each $\rho_{mn}$ can be expanded as the ratio between the determinants and the cofactors, then we have:
\begin{equation}
	I = \frac{I_1}{I_{2,1}+I_{2,2}+I_{2,3}+I_{2,4}+I_{2,5}} \,.
\end{equation}
The numerator term $I_1$ determines when the leakage current vanishes, i.e., $I_1=0$, and the denominator term $I_2$ determines whether a sharp peak will occur, i.e., $I_2=0$.
Although the leakage current can be evaluated analytically in terms of the matrix elements from Eq.~\eqref{EQ: Total Hamiltonian Matrix Form}, these matrix elements dynamically depend on the magnetic field, quantum dot geometries, and interdot distance.
Therefore, only in limited cases can a short expression for the leakage current be present in the main text, so full expressions are provided in the supplementary materials.
\begin{figure}[htbp]
    \centering
    \includegraphics[width=1\columnwidth]{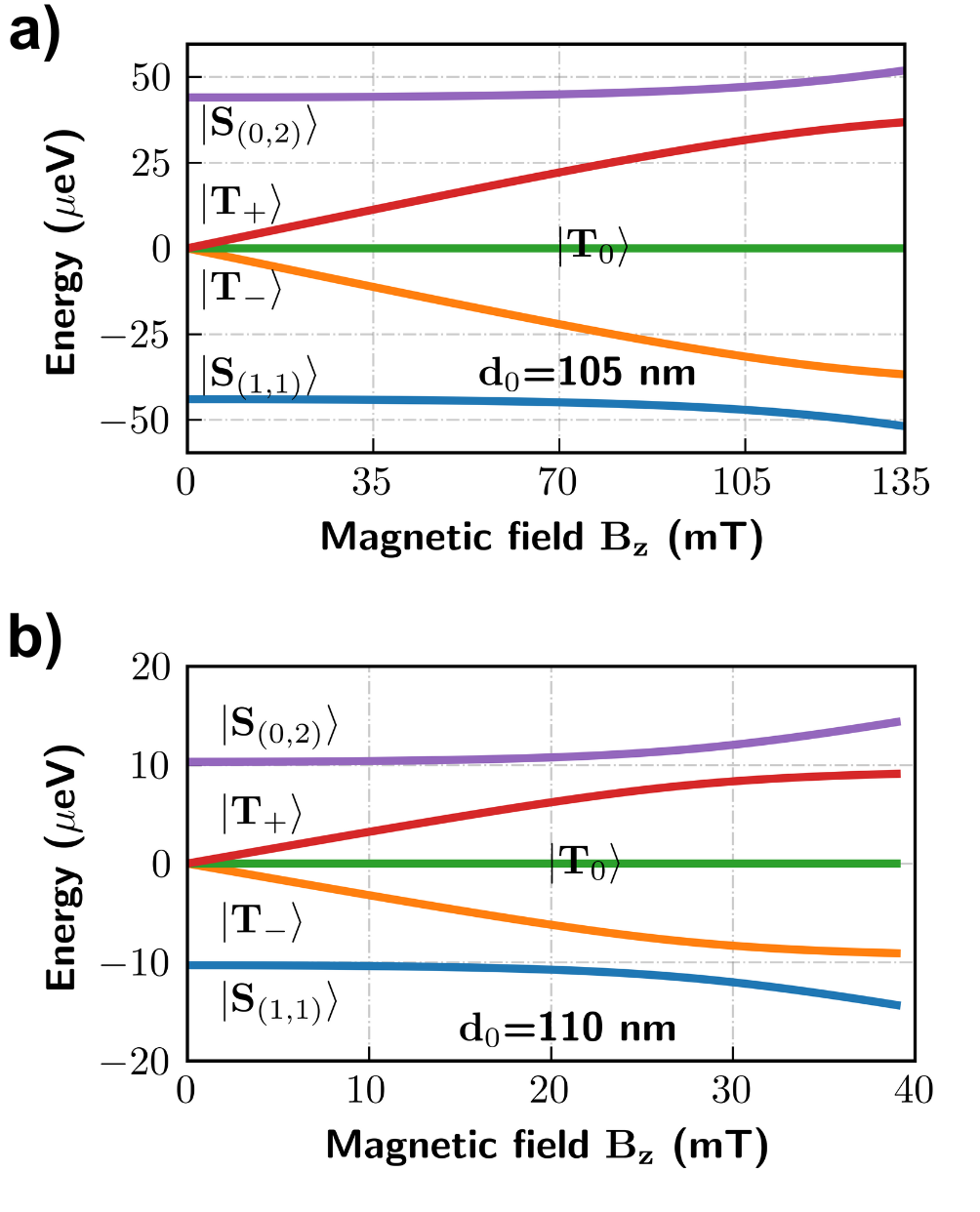}
    \caption{
    The energy levels of $H$ as a function of the out-of-plane magnetic field at $\delta=0$.
    At $B_z$=0, the energy levels for the three (1,1)-triplet states are degenerated having the same energy.
    The $\ket{S_{(1,1)}}$ state is below the $\ket{T_0}$ due to the finite exchange interaction.
    As $B_z$ increase, both the orbit magnetic field term and the Zeeman term are activated, however, the orbital magnetic field terms only have magrinal effect and are further dimmed in the range of the magnetic field we considered.
    The Zeeman terms will create a splitting between $\ket{T_+}$ and $\ket{T_-}$ of magnetude $\varepsilon_{\text{Z}}$.
    The parameters used to generate this plots are as the following: $R_{a,0}$=15~nm, $e=0.8$, $\delta=0$~meV.
    }
    \label{Fig: Eng vs Bz}
\end{figure}
\begin{table}[tbp!]
\caption{
A summary of parameters used in this work.
In this table, the out-of-plane heavy-hole band and light-hole band mass is defined as $m_{\text{HH}}=m_0/\pqty{\gamma_1-2\gamma_2}$, $m_{\text{LH}}=m_0/\pqty{\gamma_1+2\gamma_2}$; the in-plane heavy-hole band effective mass is defined as $m_{\text{HP}}=m_0/\pqty{\gamma_1+\gamma_2}$.
$m_0$ is the bare electron mass, $\gamma_1$, $\gamma_2$, $\gamma_3$ are Luttinger parameters.
}
\label{TB1 - Germanium DQD parameter}
\begin{ruledtabular}
\begin{tabular}{ll}
\textrm{Parameters}&
\textrm{Value}\\
\colrule
Luttinger parameters $\gamma_1$ & 13.38 \\
Luttinger parameters $\gamma_2$ & 4.24\\
Luttinger parameters $\gamma_3$ & 5.69  \\
Effective mass $m_{\text{LH}}$ &  0.204 $m_0$ \\
Effective mass $m_{\text{HH}}$ &  0.046 $m_0$ \\
Effective mass $m_{\text{HP}}$ & 0.057 $m_0$  \\
Quantum well width $L$ & 11 nm \\
Left dor radius $R_{a,0}$ & 15 nm \\
Aspect ration $e=R_{b,y}/R_{b,x}$ & 0.8 \\
Interdot distance $2d_0$ & 55 nm \\
Typical magnetic field $B_z$ & 10 mT \\
Cyclotron frequency $\omega_c$ & 31 GHz \\
Quatum well width & 10 nm \\
Gate electric field $F$ & 10 MV/m \\
Rashba SOC $\alpha_{R1}$ & 0.1 eV$\cdot$nm \\
\end{tabular}
\end{ruledtabular}
\end{table}

\section{Results and Discussion}
\label{Section 3 - Results and Discussions}
In this section, we present the main results, beginning with the energy levels of the total Hamiltonian as a function of the detuning $\delta$ and the out-of-plane magnetic field $B_z$.
Next, we discuss the dependence of the leakage current on the detuning, magnetic field, and quantum dot orientation.
We predict that, in an out-of-plane magnetic field, the leakage current exhibits strong anisotropy for an elliptical quantum dot, with the maximum leakage current appearing when the major semi-axis is aligned with the x-axis, and a stronger anisotropy leads to a larger leakage current.
\begin{figure}[htbp]
    \centering
    \includegraphics[width=1\columnwidth]{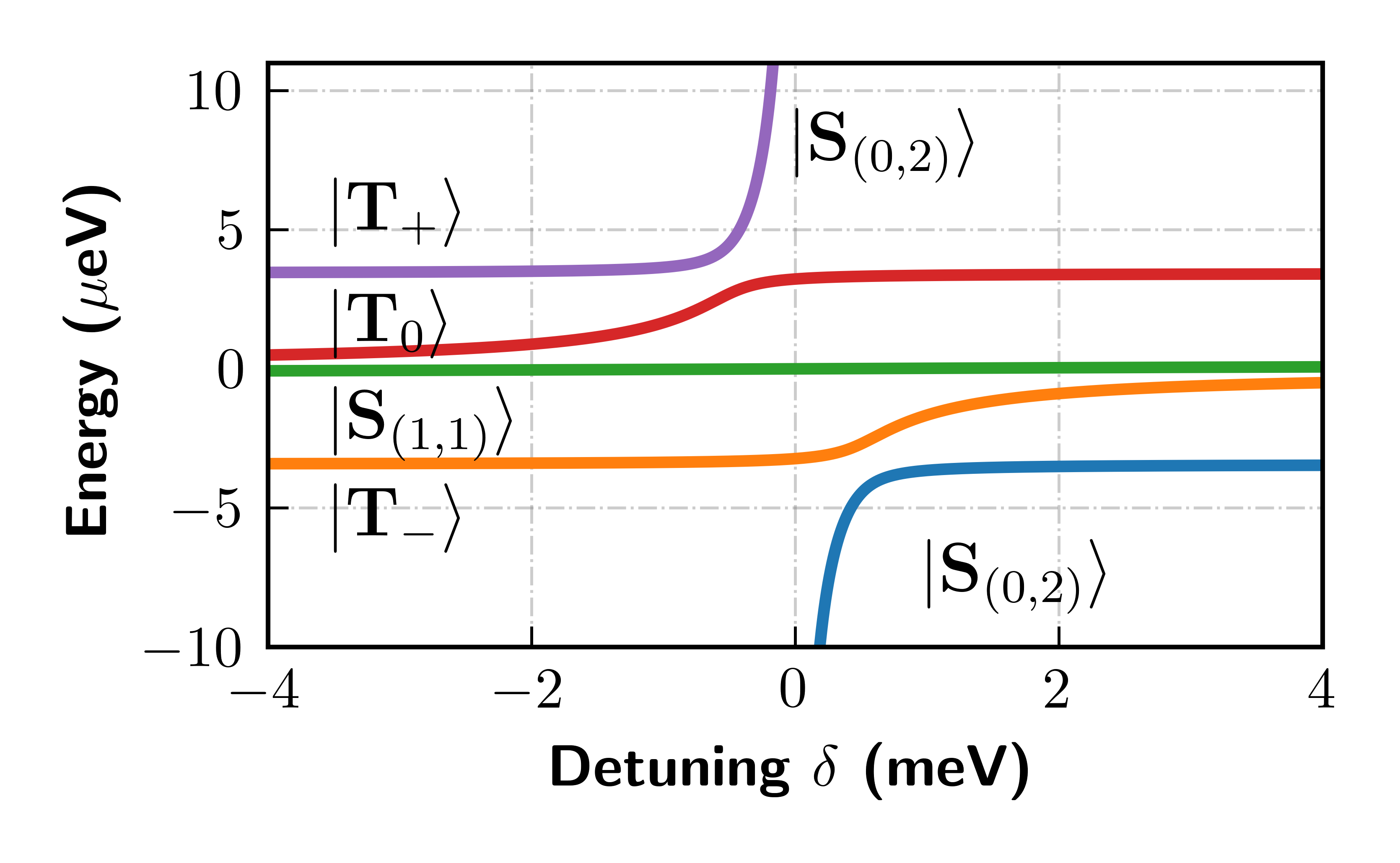}
    \caption{
    The energy levels of $H$ as a function of the detuning $\delta$.
    These energy levels are obtained by assuming the left quantum dot is circular, with a radius of $R_{a,0}$=15\,nm, and the right quantum dot is elliptical, with a semi-minor axis (along the x-axis) of $R_{b,x}$=15\,nm and an aspect ratio of $e$=0.8.
    The two quantum dots are separated by 105\,nm.
    The magnetic field is fixed at $B_z$=10\,mT, with $g_L=6.5$ and $g_R=4.5$.
    The parameters used to generate plots in the following text are selected from Ref.~\cite{Jirovec2022}.
    }
    \label{Fig: Eng vs Delta}
\end{figure}

It is worth noting that All parameters in Eq.~\eqref{EQ: Total Hamiltonian Matrix Form} are numerically calculated as outlined in Sec.~\ref{Section 2 - Subsection 1 - Model Hamiltonian} and \ref{Section 2 - Subsection 2 - Hund-Muliken Approach}, and these then serve as inputs for the analytical solution of the leakage current expression.
All the results presented in this section are based on a hybrid analytical-numerical approach, which is needed in order to account for the complex dependence of the quantum dot wave functions on the magnetic field, interdot distance, and ellipticity.

This section is organized as follows: In Sec.~\ref{Section3 - Subsection 1: Energy levels}, we discuss the energy levels associated with the $(1,0)-(0,2)$ qubit Hilbert space, identifying regions where qubit control and signal readout are implemented.
In Sec.~\ref{Section3 - Subsection 2: Leakage current}, we present the expression for the leakage current and discuss the origin of the shift and the vanishing of the leakage current, identifying regions for better determination of the SOC and for reducing the signal-to-noise ratio.

\subsection{Energy levels}
\label{Section3 - Subsection 1: Energy levels}
In investigating the energy dispersion as a function of detuning and magnetic field, we numerically diagonalize the matrix Eq.~\eqref{EQ: Total Hamiltonian Matrix Form}.
The parameters used in this section are summaized in table.~\ref{TB1 - Germanium DQD parameter}.
To understand the energy levels as a function of the detuning $\delta$ or the magnetic field $B_z$, we start from the simplest case, where $B_z$ and SOC matrix elements are not present.
In this situation, the three (1,1)-triplet states are degenerate, the (1,1)-triplet states share the same energy $E_(1,1)=0$, while the mixed $\ket{S_{(1,1)}}$ and $\ket{S_{(0,2)}}$ states has a energy gap $\sqrt{4t_{\text{C}}^2+(\delta-\Delta u)^2}$.
In the experimentally relevant magnetic field range (0-100\,mT), for an interdot distance (105\,nm), the Coulomb exchange energy $\Delta u$ is around 3\,neV, which is much smaller than the charge tunneling matrix element $t_{\text{C}}$ (around 40\,$\mu$eV).
Therefore, in this limit, the energy gap becomes to $\sqrt{4t_{\text{C}}^2+\delta^2}$.
This feature is reflected in Fig.~\ref{Fig: Eng vs Bz}, where we can notice that the gap between $\ket{S_{(0,2)}}$ curve and $\ket{S_{(1,1)}}$ curve is $2t_{\text{C}}$.
Meanwhile, the exchange coupling defined as the energy difference between the $\ket{T_0}$ and $\ket{S_{(1,1)}}$ states (indeed the $\ket{S_{(1,1)}}$-$\ket{S_{(0,2)}}$ maxing state) can be evaluated as $J=\delta/2-\sqrt{\delta^2+4t_{\text{C}}^2}/2$ as seen Ref.~\cite{Jirovec2021, Jirovec2022, Liles2024}.
At large detuning, the exchange coupling is $J=2t_{\text{C}}^2/\delta$, opening the possibility of pure electrical control of the exchange energy via $\delta$, which is a key property for enabling spin state readout \cite{Engel2004, Taylor2005}.

At a finite magnetic field, as shown in Fig.~\ref{Fig: Eng vs Bz}, the degeneracy of the three (1,1)-triplet states is lifted, primarily by the Zeeman term $\varepsilon_{\text{Z}}$ which is linear in $B_z$.
The $\ket{T_0}$ state is not affected by the Zeeman term, therefore remaining a straight line at $E=0$.
In Fig.~\ref{Fig: Eng vs Bz}a), at larger magnetic field, the linearly increasing $\ket{T_+}$ band starts to anti-cross with the $S_{(0,2)}$ band leading to a strong band mixing, with a similar result shown between $\ket{T_-}$ and $S_{(1,1)}$.
This trend is clearer in Fig.~\ref{Fig: Eng vs Bz}b), where the interdot distance is larger and the charge tunneling term $t_{\text{C}}$ is smaller, therefore, the energy gap between $S_{(0,2)}$ and $S_{(1,1)}$ are smaller, leading to the anti-crossing appearing in smaller $B_z$.

We now focus on the energy levels as a function of the detuning $\delta$ at finite magnetic field, as shown in Fig.~\ref{Fig: Eng vs Delta}.
At a large negative detuning $\delta$, the $\ket{S_{(0,2)}}$ state are far seperated from the (1,1)-states.
This situation corresponds to the standard initialization of the singlet-triplet qubit, where the large detuning allows the two carriers to be in the same quantum dot.
Further tuning of the gate bias voltage will bring the system into the $\ket{S_{(1,1)}}$ and $\ket{T_0}$ subspace.
The Zeeman term splits the $\ket{T_+}$ and $\ket{T_-}$ from $\ket{T_0}$.
As the detuning approaches to $0$, there will be a smaller energy difference between the (1,1)-states and the (0,2)-states.
The SOC and charge tunelling strongly mix the $\ket{T_0}$ and $\ket{S_{(1,1)}}$ states, resulting in a large exchange coupling.
This regime defines the working point of the singlet-triplet qubit.

\subsection{Leakage current}
\label{Section3 - Subsection 2: Leakage current}
Next, we discuss the enhancement of the leakage current as a function of the detuning $\delta$, the magnetic field $B_z$, and the quantum dot geometries.
We first consider the interdot distance $2d_0$ to be large enough that the Coulomb exchange energy $\Delta u$ can be disregarded.
The leakage current can be written as $I_1/I_2$, where the numerator takes the form:
\begin{widetext}
\begin{equation}\label{EQ: I_1 simplified}
    I_1=64 t_{\text{SO},1}^2 t_{\text{SO},2}^2 \varepsilon_{\text{Z}}^2 (4t_{\text{SO},1}^2t_{\text{SO},3}^4-t_{\text{SO},3}^2(2(t_{\text{SO},1}^2+t_{\text{SO},2}^2)+\varepsilon_Z^2)t_{\text{C}}^2+t_{\text{SO},2}^2t_{\text{C}}^4)^2 \,.
\end{equation}
\end{widetext}
The denominator $I_2$ consists of five terms, making it lengthy for the main text.
Therefore, the complete expression is provided in supplementary material.
From Eq.~\eqref{EQ: I_1 simplified}, we can identify four situations that a leakage current will vanish: $\varepsilon_{\text{Z}}=0$, $t_{\text{SO},1}=0$, $t_{\text{SO},2}=0$, or the quantum dot gometries and out-of-plane magnetic field just allows the last term in Eq.~\eqref{EQ: I_1 simplified} vanish.
Additionally, there is a trivial situation where the detuning is large enough that the (1,1)-states are far away from the (0,2)-states, leading to a vanishing leakage current, as represented in the large detuning region in Fig.~\ref{Fig: Eng vs Delta}.

We notice that similar results have been reported in an earlier work in Ref.~\cite{Qvist2022_2}, where several stopping points of the leakage current is discussed, due to an interplay between the vector external magnetic field and and a general spin-orbit coupling vector (the z-component is also incorperated).
While Ref.~\cite{Qvist2022_2} considered a broader aspect of leakage current, we focus on the impact of the geometric design of double quantum dot systems, specifically the aspect ratio of individual quantum dots and the case where the two quantum dots are not aligned.

In the simplest case, two identical circular quantum dots are considered (with an aspect ratio of $e=1$), which represents an ideal scenario from the point of view of quantum dot fabrication.
Regardless of the quantum dot radius or the interdot distance, the molecular orbital wave functions for the left and right quantum dots will have the same shape.
As a result, the matrix elements $\mel**{L}{H_{\text{SO}}}{L} + \mel**{R}{H_{\text{SO}}}{R}$ will sum to zero, meaning that $t_{\text{SO},1}=0$, leading to a vanishing leakage current, as indicated in Eq.~\eqref{EQ: I_1 simplified}.
\begin{figure}[!ht]
    \centering
    \includegraphics[width=1\columnwidth]{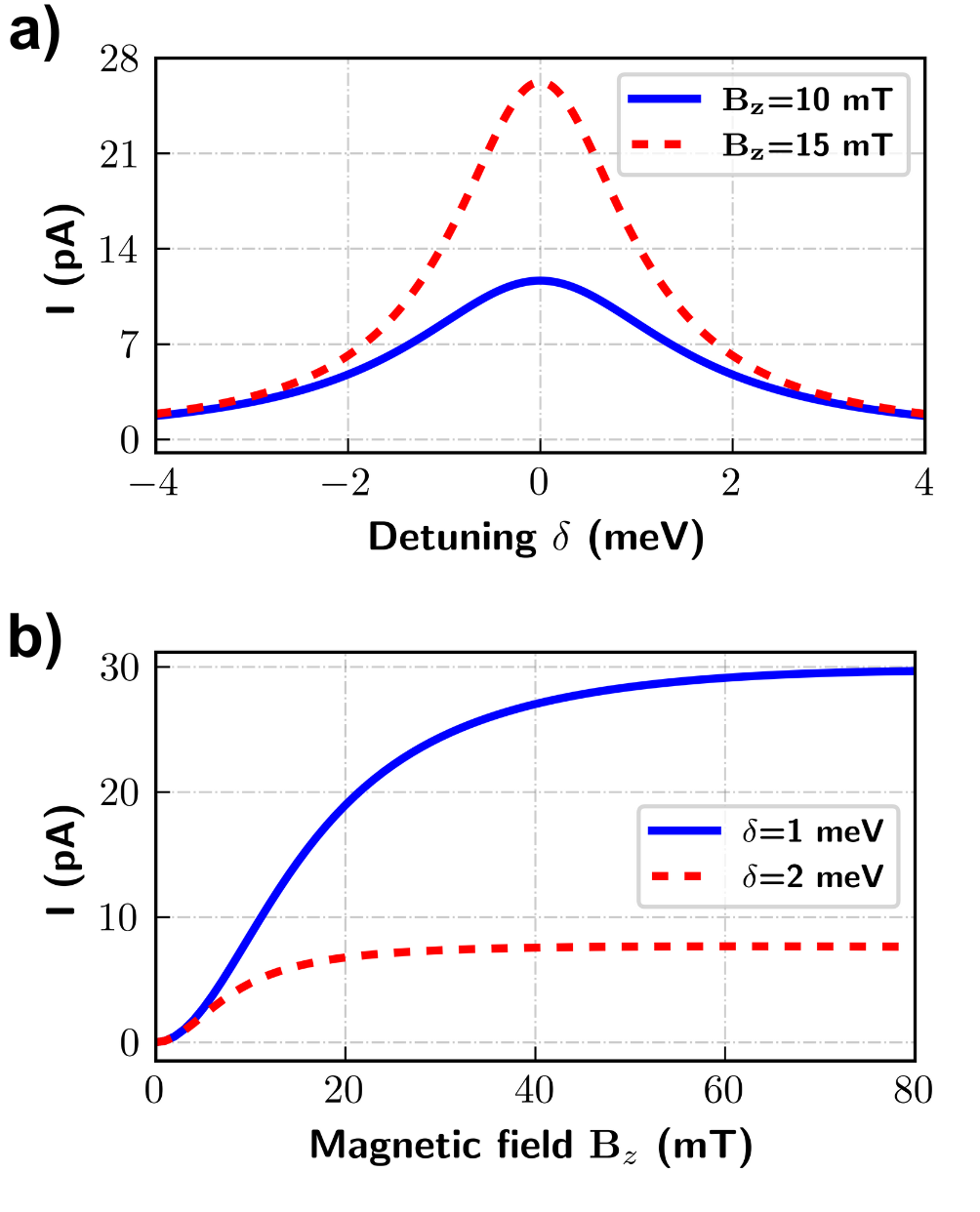}
    \caption{
    The leakage currents as the function of the detuning $\delta$ and the magnitude of the out-of-plane magnetic field $B_z$.
    a) The leakage current as a function of the detuning $\delta$ in different magnetic fields. The solid blue curve is for $B_z$=15\,mT case, where the dotted red curve is for $B_z$=10\,mT case.
    b) The leakage current as a function of the magnetic field in different detunings.
    In both plots, the left quantum dot is circular with a radius of $R_{a,0}$=15\,nm, and the right quantum dot is elliptical with a semi-minor axis (along the x-axis) of $R_{b,x}$=15\,nm and an aspect ratio of $e$=0.8, the interdot distance is $2d_0$=105\,nm.
    }
    \label{Fig: I Total 105nm}
\end{figure}

We next discuss the leakage current $I$ when the right quantum dot is elliptical, starting with its dependence on the detuning parameter $\delta$.
Fig.~\ref{Fig: I Total 105nm} shows the leakage current as a function of the detuning and the out-of-plane magnetic fields.
{In Fig.~\ref{Fig: I Total 105nm}a), we notice that there is a unique maximum as a function of the detuning, indicating a balance between the (1,1)-(2,0) state mixings.
When the magnetic field is small, both the Zeeman term and orbital magnetic field term are weak, $\delta_{\text{max}}$ will not shift and is given by:
\begin{equation}
  \label{EQ: Delta Max}
  \delta_{\text{max}}=\frac{2 t_{\text{C}} \left(2 t_{\text{SO},1}^2-t_{\text{C}}^2\right) t_{\text{SO},2} t_{\text{SO},3}}{t_{\text{C}}^2 t_{\text{SO},2} + 2 t_{\text{SO},1}^2 t^2_{\text{SO},3}} \,.
\end{equation}
As the detuning diverge from $\delta_{max}$, the system will be tuned to the (2,0)-subspace, where the leakage current will vanish.}
In Fig.~\ref{Fig: I Total 105nm}b), we plot the leakage current as a function of the out-of-plane magnetic field in different detunings.
{At $B_z=0$, the time-reversal symmetry is recovered, the Zeeman term vanishes and according to Eq.~\eqref{EQ: I_1 simplified}, the leakage current will vanish.
As the magnetic field increases, the (1,1)-triplet states are splitted, reducing the gap between the $\ket{T_+}$ and $\ket{S_{(0,2)}}$ states.
Therefore, the leakage current will firstly increase then saturate to a maximum, then gradually decrease once the gap is enlarged again.
During this process, a larger magnetic field will shrick the quantum dot size, however, in the regime of the magnitude we considered here, this effect is marginal.}
\begin{figure}[!ht]
    \centering
    \includegraphics[width=1\columnwidth]{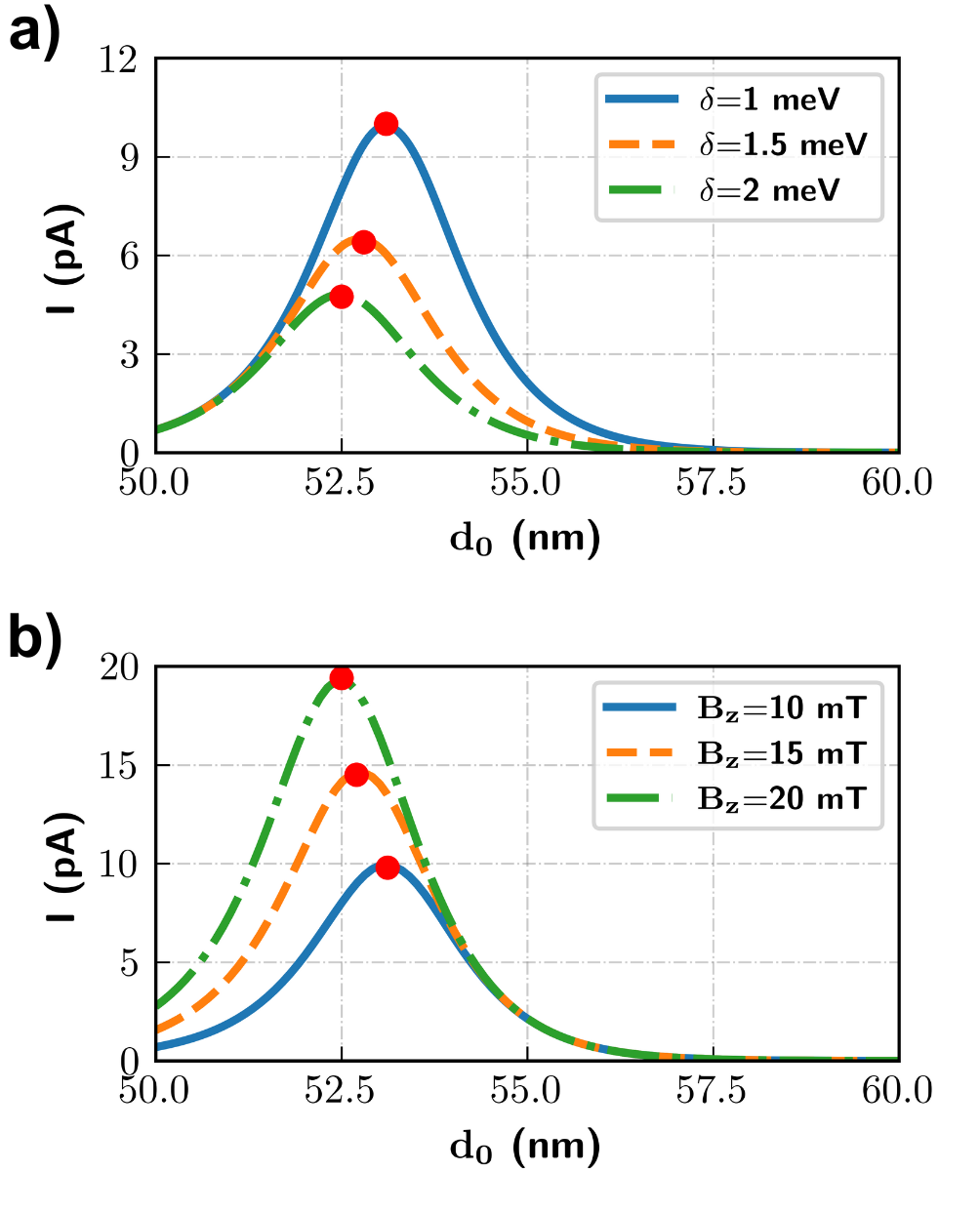}
    \caption{
    The leakage currents as the function of $d_0$.
    The interdot distance is $2d_0$.
    a) The leakage current as a function of $d_0$ in different detuning.
    b) The leakage current as a function of $d_0$ in different magnetic fields.
    {
    In both plots, the left quantum dot is circular with a radius of $R_{a,0}$=15\,nm, and the right quantum dot is elliptical with a semi-minor axis (along the x-axis) of $R_{b,x}$=15\,nm and an aspect ratio of $e$=0.8.}
    }
    \label{Fig: I vs d0 Total}
\end{figure}

Next, we study the leakage current as a function of the interdot distance.
In Fig.~\ref{Fig: I vs d0 Total}, we plot the leakage current as a function of the interdot distance $d_0$ in different detunings and magnetic fields.
When the interdot distance is small, $t_{\text{C}}$ is large therefore, the SOC matrixs can not overcome the the gap $2t_{\text{C}}$ as discussed in \ref{Section3 - Subsection 1: Energy levels}, therefore the leakage current will be small.
As the interdot distance increases, the wavefunctions of the holes will overlap less thus expericing local deformation potentials and electric field gradient, therefore while the tunneling matrix elements $t_{\text{C}}$ and $t_{\text{SO},3}$ decrease, the on-site SOC matrix elements $t_{\text{SO},1}$ and $t_{\text{SO},2}$ is getting stronger, leading to an increase of the leakage current.
Again, as the interdot distance further increases, the overlap of two wave functions will be negligble, and the tunelling channel will be closed, leading to a vanishing leakage current as shown in Fig.~\ref{Fig: I vs d0 Total}.
As a comparison, we plot the leakage current as a function of the interdot distance in different detunnings and magnetic fields, as expected in Fig.~\ref{Fig: I Total 105nm}.
\begin{figure}[!ht]
    \centering
    \includegraphics[width=1\columnwidth]{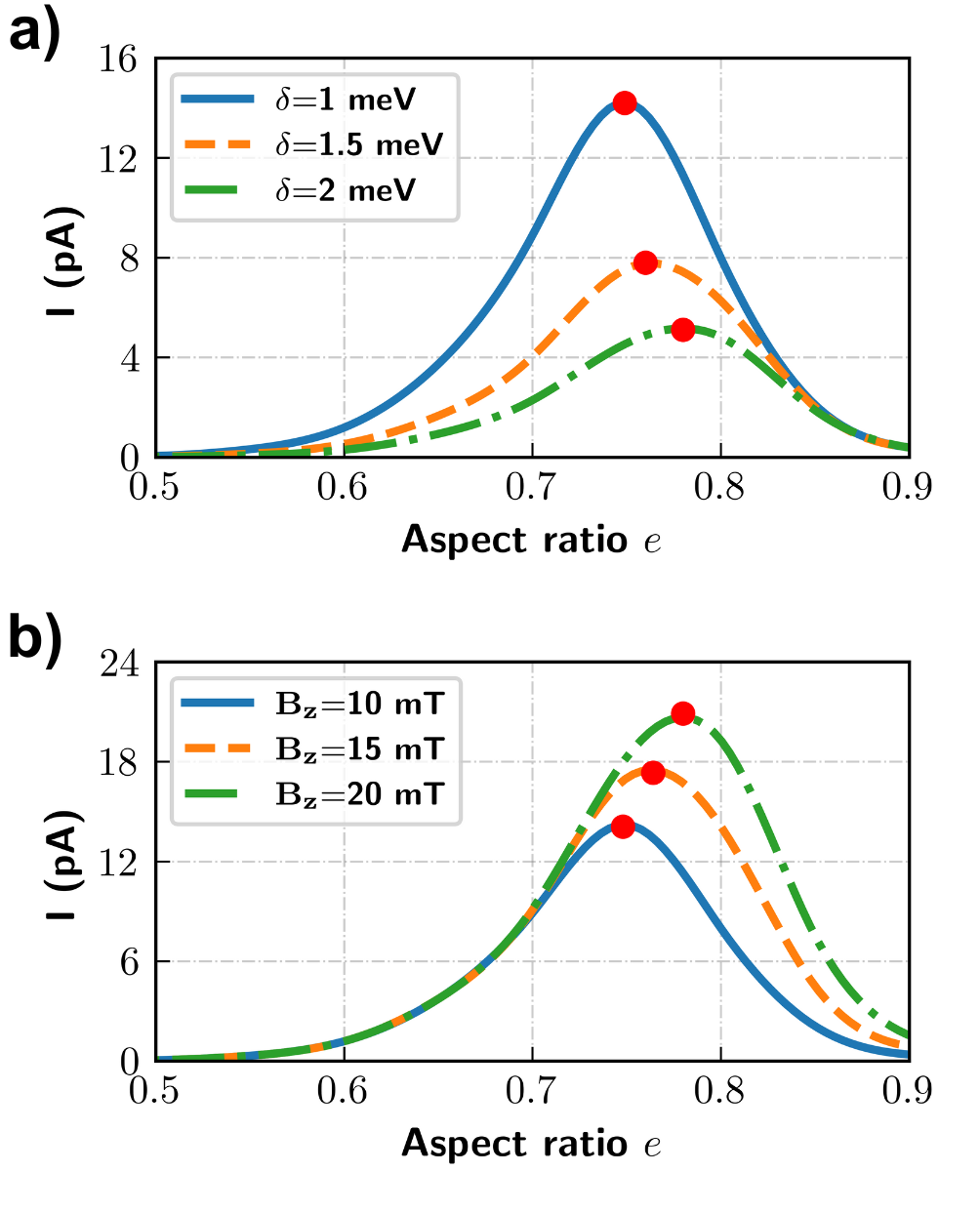}
    \caption{
    The leakage currents as the function of the aspect ratio $e$.
    $e$ is defined as $R_{b,y}=R_{b,x}/e$.
    An aspect ratio approaching to $1$ indicates the right quantum dot converge to a circular quantum dot.
    a) The leakage current as a function of $e$ in different detunings.
    b) The leakage current as a function of $e$ in different magnetic fields.
    In both plots, the left quantum dot is circular with a radius of $R_{a,0}$=15\,nm, and the right quantum dot is elliptical with a semi-minor axis (along the x-axis) of $R_{b,x}$=15\,nm and an aspect ratio of $e$=0.8, the interdot distance is $2d_0$=105\,nm.}
    \label{Fig: I vs Ellipticity Total}
\end{figure}

In Fig.~\ref{Fig: I vs Ellipticity Total}, we present the leakage current as a function of the aspect ratio $e$.
A smaller aspect ratio indicates a strong anisotropy of the right quantum dot, when the aspect ratio satisfies $e=1$, the two quantum dots are identically circular.
As the aspect ratio approaches 1, the symmetry of the wave-functions is restored, the matrix elements $t_{\text{SO},1}$ will approach to zero, leading to a vanishing leakage current, which is reflected in the Fig.~\ref{Fig: I vs Ellipticity Total}.
When the aspect ratio is too small, the right quantum dot will be squeezed along the $x$-axis, leading to a smaller wave function overlap, therefore the leakage current will also decrease.
Only between the two extremes, there exists a maximum leakage current as a function of the aspect ratio, as a result of the complicated interplay between the SOC matrix elements and the charge tunneling matrix element.
In general, considering the leakage current at different values of the detuning and magnetic field strength, we observe that although the overall shapes of the leakage current curves remain similar, their magnitudes exhibit a clear decreasing trend with increasing detuning.
Likewise, as the magnetic field strength increases, the leakage current further diminishes, a behavior that can be attributed to the trends illustrated in Fig.~\ref{Fig: I Total 105nm}.
\begin{figure}[tbp]
    \centering
    \includegraphics[width=1\columnwidth]{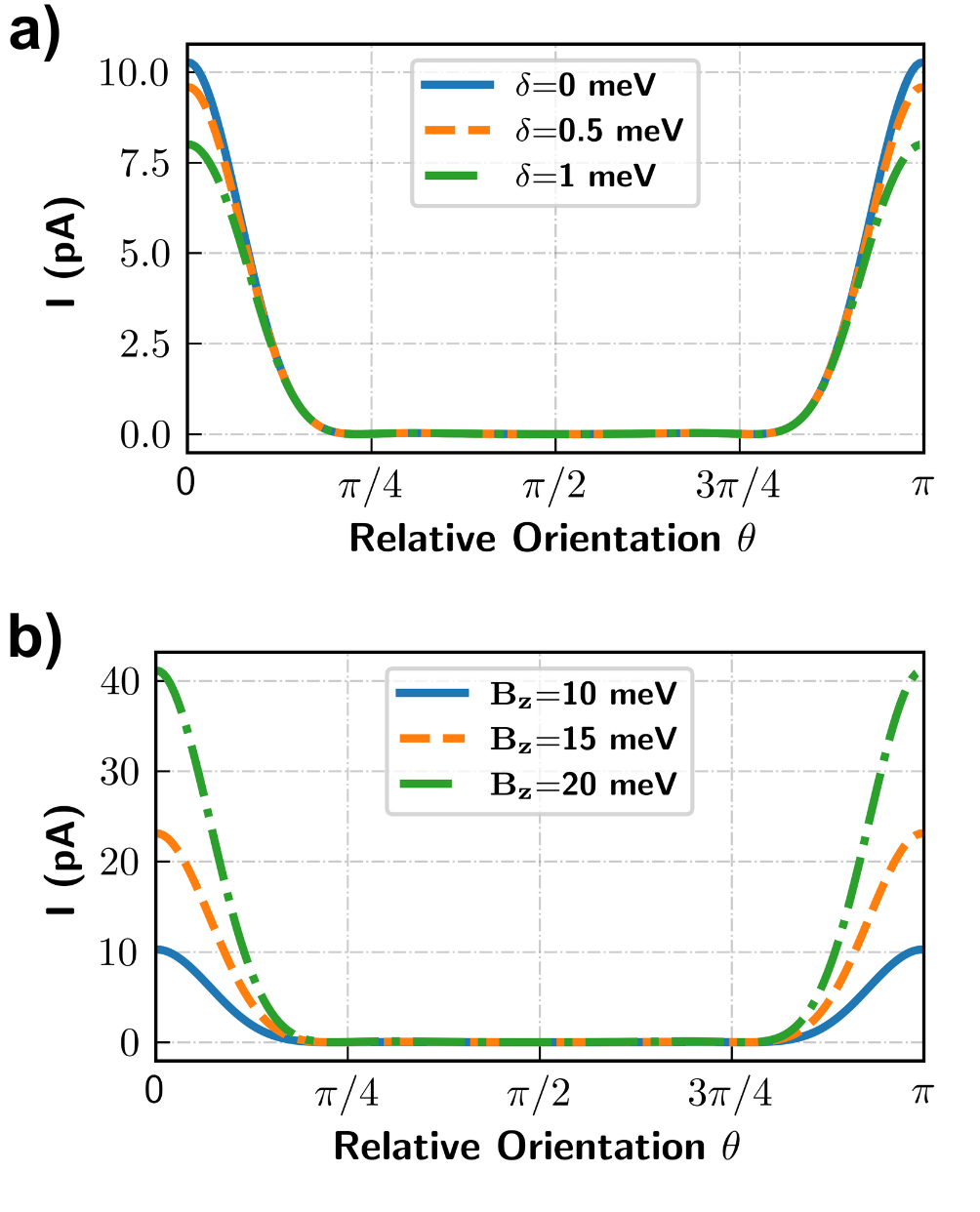}
    \caption{
    The leakage currents as a function of the relative orientation angle $\theta$ (the angle between the principal axis and the x-axis) of the right elliptical quantum dot.
    a) The leakage current as a function of $\theta$ in different detunings.
    b) The leakage current as a function of $e$ in different magnetic fields.
    In both plots, the left quantum dot is circular with a radius of $R_{a,0}$=15\,nm, and the right quantum dot is elliptical with a semi-minor axis (along the x-axis) of $R_{b,x}$=15\,nm and an aspect ratio of $e$=0.8, the interdot distance is $2d_0$=105\,nm.
    }
    \label{Fig: I vs Theta Total}
\end{figure}

Finally, we study the leakage current when the principal axis of the right quantum dot has a finite angle with respect to the x-axis, while the aspect ratio remains fixed.
This simulates cases where the principal axis of the quantum dots is not aligned with the lab frame, which are relevant to experiments such as those in Ref.~\cite{Jirovec2022, Liles2024}.
The main results are presented in Fig.~\ref{Fig: I vs Theta Total}, which shows the leakage current as a function of the relative orientation angle $\theta$ (counterclock-wise) of the right elliptical quantum dot, ranging from $0$ to $\pi$.
We observe that the leakage current reaches its maximum at $\theta = 0$ (and $\theta = \pi$), indicating that the largest leakage occurs when the two major semi-axes are parallel.
When the angle increases, the leakage current decreases rapidly, primarily due to the resonant-type behavior of $I_1$ and $I_2$ described in Eq.~\eqref{EQ: I_1 simplified}.
Moreover, there exists a broad range of relative orientation angles where the leakage current becomes vanishingly small, with symmetry about $\theta = \pi/2$.
As the relative orientation angle approaches $\pi/2$, the overlap of the wave functions of the two quantum dots increases.
As a result, the charge tunneling matrix elements $t_{\text{C}}$ rise significantly, dominating the denominator of the leakage current expression and leading to a rapid decay in the leakage current.

In Fig.~\ref{Fig: I vs Theta Total}a) we investigate the leakage current as a function of $\theta$ under different detuning magnitudes.
Consistent with the results above, the detuning appears only in the $I_2$ term and does not affect other matrix elements.
As a result, we observe a family of curves with varying maximum leakage currents but with same trends. Next, Fig.~\ref{Fig: I vs Theta Total}b) shows the leakage current as a function of $\theta$ for different magnetic field magnitudes.
In a similar manner to Fig.~\ref{Fig: I vs Theta Total}a), the leakage current rapidly decays as $\theta$ approaches $\pi/2$.
Additionally, as the magnetic field increases, the leakage current also increases.
The behavior of the leakage current can be better understood quantitatively by examining the variations of the matrix elements in Eq.~\eqref{EQ: Total Hamiltonian Matrix Form} as a function of $\theta$.
When the elliptical quantum dot rotates, the overlap integral $S$ will oscillate with period $\pi$, the overlap will be maximized if the major semi-axis of the right elliptical quantum dot is parallel to x-axis.
As an estimate of the variation of the overlap integral $S$, if we consider $e=0.8$ and $2d_0=105$\,nm, $S$ at $\theta=\pi/2$ is $20$ times larger then the $S$ at $\theta=0$.
Correspondingly, the $t_{\text{C}}$ term will change in a smilar way with $S$, reaching a maxima when $\theta=\pi/2$.
One new term will be activated in this case, which is the Coulomb exchange energy $\Delta u$.
It is negligible in previous discussions, however, it now can change siginificantly due to the large change in $S$.
The $\Delta u$ term can vary from $2$\,neV to $400$\,neV, and the $t_{\text{C}}$ term varies from $40$\,$\mu$eV to $600$\,$\mu$eV.
Another tuenlling term is $t_{\text{SO},3}$, which follows the trend with $t_{\text{C}}$.
The matrix elements $t_{\text{SO},1}$ and $t_{\text{SO},2}$ also have strong anisotropy, except due to the overlap integral $S$, the Rashba SOC also exihibits strong anisotropy which is reflected in the coefficients $\alpha_2$ and $\alpha_3$.
For a circular quantum dot, the $\alpha_2 \propto (\gamma_2+\gamma_3)/2 $ term will vanish while the term $\alpha_3 \propto (\gamma_2-\gamma_3)/2$ is dominating.
For an elliptical quantum dot, the $\alpha_2$-Rashba SOC term will be activated and can contribute equivalently to $t_{\text{SO},1}$ or $t_{\text{SO},2}$ as the $\alpha_3$-Rashba SOC term \cite{Abhik2023}.
Our numerical results suggest that the magnitude of $\mel**{R}{H_{\text{SO}}}{R}$ term has a minima at $\theta=\pi/2$ with period $\pi$ leading to a minima in $t_{\text{SO},2}$.
The variations of the magnitude of the matrix elements are presented in the supplementary material.

The methods developed in this work provide a connection between the leakage current and controllable parameters, such as in-plane confinement, quantum dot anisotropy, magnetic fields, and detuning, which can guide the investigation of PSB in various platforms.
We note that our method focuses on single-hole occupation, while higher orbital excited states are not taken into account. Extending our treatment to include in-plane magnetic fields requires adding both an off-diagonal Zeeman term ($g_{\parallel}\mu_BBx$) to Eq.~\eqref{EQ: Total Hamiltonian Matrix Form} and the appropriate vector potentials in the gauge term, with the quantum wave functions determined numerically.
Increasing the numerical accuracy would require including many excited states to solve the individual quantum dot wave functions, as orbital magnetic field terms become significant — this is beyond the scope of the current work. Similarly, inhomogeneous strain and shear strain have proven to be important in modifying $g$-factors and SOC in single quantum dot hole spin qubits \cite{Liles2021, Venitucci2020, Carlos2022, Mena2023}. Therefore, we expect that strain and its gradients can also be used to improve the leakage current and possibly enhance the operation time of singlet-triplet qubits in group IV semiconductor quantum dot spin qubits. These possibilities will be explored further in future projects.

\section{Summary and Outlook}
\label{Section 4 - Conclusion and Outlook}
We have studied the leakage current in a two-hole double quantum dot configuration in the PSB regime, based on a germanium platform, where one of the quantum dots is elliptical and free to rotate, inspired by current experiments \cite{Jirovec2022, Liles2024}.
Using a $k^3$-Rashb, $k$-linear Rashba SOC model and analytically solving the Lindblad equation, we derive expressions for the leakage current under various conditions.
Using a hybrid numerical-analytical method, we find that there exists a specific detuning $\delta_{\text{max}}$ in Eq.~\eqref{EQ: Delta Max}, where the leakage current is maximized.
In all situations studied the leakage current decreases with increasing interdot distance and stronger in-plane magnetic fields.
The leakage current is strongly anisotropic and increases significantly when the semi-major axis is aligned with the $x$-axis.
Our results can serve as a probe to study the site-dependent $g$-factors and SOC, and to enhance the leakage current by optimizing quantum dot geometries, detuning, and magnetic field orientation.
Although we limit our focus to out-of-plane magnetic fields and the germanium heterostructure platform in this work, the formalism can be extended to silicon metal-oxide-semiconductor devices by performing a full numerical diagonalization of the total Hamiltonian $H$, as the perturbative treatment of the SOC is less effective \cite{Wang2024}.

\acknowledgments
We thank Ik Kyeong Jin and Robert Joynt for enlightening discussions.
ARH acknowledges ARC funding from grant IL230100072, DP200100147 and DP210101608.
D.C. acknowledges support from the Australian Research Council (ARC) through Future Fellowship FT190100062 and Discovery Project DP2401062.
\clearpage
\renewcommand{\theequation}{\thesection\arabic{equation}}
\renewcommand{\thefigure}{\thesection\arabic{figure}}
\onecolumngrid
\appendix
\section{Expressions for leakage current}
In this section, we present the full expression for the leakage current in the limit $\Delta u$=0 at large enough interdot distance.
The numerator part of the leakage current is:
\begin{equation}
    I_1=64 t_{\text{SO},1}^2 t_{\text{SO},2}^2 \varepsilon_{\text{Z}}^2 (4t_{\text{SO},1}^2t_{\text{SO},3}^4-t_{\text{SO},3}^2(2(t_{\text{SO},1}^2+t_{\text{SO},2}^2)+\varepsilon_Z^2)t_{\text{C}}^2+t_{\text{SO},2}^2t_{\text{C}}^4)^2 \,.
\end{equation}
The denominator part can be written as
\begin{equation}
    I_2 = I_{2,1} + I_{2,2} + I_{2,3} + I_{2,4} + I_{2,5} \,,
\end{equation}
The expression for $D$ terms are:
\begin{equation}
    I_{2,1} = 4 E_{\text{Z}}^8 t_{\text{C}}^4 t_{\text{SO},3}^2 \left( 8 t_{\text{SO},1}^2 t_{\text{SO},2}^2 + t_{\text{C}}^2 t_{\text{SO},3}^2 \right)
\end{equation}
\begin{small}
\begin{equation}
\begin{aligned}
    I_{2,2}=& 16 t_{\text{SO},1}^2 t_{\text{SO},2}^2 \left( t_{\text{C}}^2 - 2 t_{\text{SO},3}^2 \right)^2 \times \\
    &[ 4 t_{\text{C}}^6 t_{\text{SO},1}^2 t_{\text{SO},2}^2 + 8 t_{\text{C}}^4 \left( 4 t_{\text{SO},2}^4 t_{\text{SO},3}^2 + t_{\text{SO},1}^4 \left( -2 t_{\text{SO},2}^2 + t_{\text{SO},3}^2 \right) + 
2 t_{\text{SO},1}^2 \left( t_{\text{SO},2}^4 - 3 t_{\text{SO},2}^2 t_{\text{SO},3}^2 \right) \right) \\
&\quad + 32 t_{\text{C}}^3 t_{\text{SO},2} \left( t_{\text{SO},1}^2 - t_{\text{SO},2}^2 \right)^2 t_{\text{SO},3} \Delta u - 
64 t_{\text{C}} t_{\text{SO},2} \left( t_{\text{SO},1}^3 - t_{\text{SO},1} t_{\text{SO},2}^2 \right)^2 t_{\text{SO},3} \Delta u \\
& \quad + 
4 t_{\text{SO},1}^2 t_{\text{SO},3}^2 ( 8 t_{\text{SO},1}^2 \left( -t_{\text{SO},1}^2 + t_{\text{SO},2}^2 + t_{\text{SO},3}^2 \right)^2 + 
4 \left( t_{\text{SO},1}^2 - t_{\text{SO},2}^2 \right)^2 \Delta u^2 + \left( t_{\text{SO},1}^2 - t_{\text{SO},2}^2 \right)^2 \Gamma^2 \hbar^2 ) \\
& \quad + 2 t_{\text{C}}^2 \left( 8 t_{\text{SO},1}^2 \left( \left( -t_{\text{SO},1}^2 t_{\text{SO},2} + t_{\text{SO},2}^3 \right)^2 - 
2 \left( t_{\text{SO},1}^4 - 4 t_{\text{SO},1}^2 t_{\text{SO},2}^2 + 3 t_{\text{SO},2}^4 \right) t_{\text{SO},3}^2 + \left( 2 t_{\text{SO},1}^2 - 
3 t_{\text{SO},2}^2 \right) t_{\text{SO},3}^4 \right) \right.\\
& \quad + 4 t_{\text{SO},2}^2 \left( -t_{\text{SO},1}^2 + t_{\text{SO},2}^2 \right)^2 \Delta u^2 + 
t_{\text{SO},2}^2 \left( -t_{\text{SO},1}^2 + t_{\text{SO},2}^2 \right)^2 \Gamma^2 \hbar^2 ]
\end{aligned}
\end{equation}
\end{small}
\begin{equation}
\begin{aligned}
I_{2,3}=-2 E_{\text{Z}}^6 t_{\text{C}}^2 \times &[ 4 t_{\text{C}}^6 t_{\text{SO},2}^2 t_{\text{SO},3}^2 - 
4 t_{\text{C}}^4 \left( 2 t_{\text{SO},2}^2 t_{\text{SO},3}^4 + 
t_{\text{SO},1}^2 \left( 4 t_{\text{SO},2}^4 - 8 t_{\text{SO},2}^2 t_{\text{SO},3}^2 + 3 t_{\text{SO},3}^4 \right) \right) \\
&\quad + 64 t_{\text{C}}^3 t_{\text{SO},1}^2 t_{\text{SO},2}^3 t_{\text{SO},3} \Delta u - 
64 t_{\text{C}} t_{\text{SO},1}^2 t_{\text{SO},2} t_{\text{SO},3}^5 \Delta u + 
8 t_{\text{SO},1}^2 t_{\text{SO},2}^2 t_{\text{SO},3}^4 \left( 16 t_{\text{SO},1}^2 - 
4 \Delta u^2 - \Gamma^2 \hbar^2 \right) \\
&\quad - 4 t_{\text{C}}^2 t_{\text{SO},1}^2 t_{\text{SO},3}^2 \left( 32 t_{\text{SO},1}^2 t_{\text{SO},2}^2 + 24 t_{\text{SO},2}^4 + 6 t_{\text{SO},3}^4 + 
t_{\text{SO},2}^2 \left( 8 t_{\text{SO},3}^2 + 4 \Delta u^2 + \Gamma^2 \hbar^2 \right) \right) ]
\end{aligned}
\end{equation}
\begin{equation}
\begin{aligned}
I_{2,4}=4 E_{\text{Z}}^4 &[ t_{\text{C}}^{10} t_{\text{SO},2}^4 - 4 t_{\text{C}}^8 t_{\text{SO},2}^4 (4 t_{\text{SO},1}^2 + t_{\text{SO},3}^2) + 
32 t_{\text{C}}^7 t_{\text{SO},1}^2 t_{\text{SO},2}^3 t_{\text{SO},3} \Delta u \\
& \quad + 128 t_{\text{C}} t_{\text{SO},1}^4 t_{\text{SO},2} t_{\text{SO},3}^5 (-2 t_{\text{SO},2}^2 + t_{\text{SO},3}^2) \Delta u - 
32 t_{\text{C}}^5 t_{\text{SO},1}^2 t_{\text{SO},2}^3 t_{\text{SO},3} (6 t_{\text{SO},1}^2 + 4 t_{\text{SO},2}^2 + 
t_{\text{SO},3}^2) \Delta u \\
&\quad + 128 t_{\text{C}}^3 t_{\text{SO},1}^2 t_{\text{SO},2} t_{\text{SO},3}^3 ( 2 t_{\text{SO},2}^2 t_{\text{SO},3}^2 + 
t_{\text{SO},1}^2 (2 t_{\text{SO},2}^2 + t_{\text{SO},3}^2) ) \Delta u \\
&\quad + 16 t_{\text{SO},1}^4 t_{\text{SO},2}^2 t_{\text{SO},3}^6 \left( 8 t_{\text{SO},1}^2 + 
4 \Delta u^2 + \Gamma^2 \hbar^2 \right) \\
& \quad + 24 t_{\text{C}}^2 t_{\text{SO},3}^4 ( 
2 t_{\text{SO},1}^4 \left( -4 t_{\text{SO},2}^4 - 8 t_{\text{SO},2}^2 t_{\text{SO},3}^2 + t_{\text{SO},3}^4 \right) + 
t_{\text{SO},1}^2 t_{\text{SO},2}^4 \left( 4 \Delta u^2 + \Gamma^2 \hbar^2 \right) -16 t_{\text{SO},1}^6 t_{\text{SO},2}^2 ) \\
& \quad + 4t_{\text{C}}^6 t_{\text{SO},2}^4 t_{\text{SO},3}^4 + 
12 t_{\text{C}}^6 t_{\text{SO},1}^4 ( 4 t_{\text{SO},2}^4 - 8 t_{\text{SO},2}^2 t_{\text{SO},3}^2 + t_{\text{SO},3}^4 ) \\
& \quad +2 t_{\text{C}}^6 
t_{\text{SO},1}^2 ( 16 t_{\text{SO},2}^6 + 8 t_{\text{SO},2}^2 t_{\text{SO},3}^4 + 
t_{\text{SO},2}^4 \left( -64 t_{\text{SO},3}^2 + 
4 \Delta u^2 + \Gamma^2 \hbar^2 \right) ) \\
& \quad + 4 t_{\text{C}}^4 t_{\text{SO},1}^2 t_{\text{SO},3}^248 t_{\text{SO},1}^4 t_{\text{SO},2}^2 + 
4 t_{\text{C}}^4 t_{\text{SO},1}^2 t_{\text{SO},3}^28 ( 3 t_{\text{SO},2}^6 + 2 t_{\text{SO},2}^4 t_{\text{SO},3}^2 + 6 t_{\text{SO},2}^2 t_{\text{SO},3}^4 ) \\
& \quad +4 t_{\text{C}}^4 t_{\text{SO},1}^2 t_{\text{SO},3}^2 
t_{\text{SO},1}^2 ( 16 t_{\text{SO},2}^4 + 12 t_{\text{SO},3}^4 + 
t_{\text{SO},2}^2 ( 32 t_{\text{SO},3}^2 + 12 \Delta u^2 + 
3 \Gamma^2 \hbar^2 ) ) ]
\end{aligned}
\end{equation}
\begin{small}
\begin{equation}
\begin{aligned}
I_{2,5}=&8 E_{\text{Z}}^2 t_{\text{SO},1}^2 \times\\
& [5 t_{\text{C}}^{10} t_{\text{SO},2}^4 + 
2 t_{\text{C}}^8 t_{\text{SO},2}^2 \left( 4 t_{\text{SO},2}^4 - t_{\text{SO},2}^2 t_{\text{SO},3}^2 + 
t_{\text{SO},1}^2 (-16 t_{\text{SO},2}^2 + 6 t_{\text{SO},3}^2) \right) \\
& \quad +32 t_{\text{C}}^7 t_{\text{SO},2}^3 (2 t_{\text{SO},1}^2 + t_{\text{SO},2}^2) t_{\text{SO},3} \Delta u - 
128 t_{\text{C}} t_{\text{SO},1}^2 t_{\text{SO},2} (t_{\text{SO},1}^2 + t_{\text{SO},2}^2) t_{\text{SO},3}^5 (4 t_{\text{SO},2}^2 - 
t_{\text{SO},3}^2) \Delta u \\
&\quad - 32 t_{\text{C}}^5 t_{\text{SO},2}^3 t_{\text{SO},3} (6 t_{\text{SO},1}^4 + 2 t_{\text{SO},2}^4 + 
5 (t_{\text{SO},1}^2 + t_{\text{SO},2}^2) t_{\text{SO},3}^2) \Delta u \\
& \quad + 64 t_{\text{C}}^3 t_{\text{SO},2} t_{\text{SO},3}^3 \left( 8 t_{\text{SO},1}^2 t_{\text{SO},2}^2 (t_{\text{SO},1}^2 + t_{\text{SO},2}^2) + (t_{\text{SO},1}^4 - 
4 t_{\text{SO},1}^2 t_{\text{SO},2}^2 + 5 t_{\text{SO},2}^4) t_{\text{SO},3}^2 \right) \Delta u \\
&\quad + 2 t_{\text{C}}^6 \left( 8 t_{\text{SO},2}^8 - 66 t_{\text{SO},2}^4 t_{\text{SO},3}^4 2 t_{\text{SO},1}^4 \left( 12 t_{\text{SO},2}^4 - 24 t_{\text{SO},2}^2 t_{\text{SO},3}^2 + t_{\text{SO},3}^4 \right)\right) \\
&\quad +2 t_{\text{C}}^6 \left( 
t_{\text{SO},2}^6 \left( -48 t_{\text{SO},3}^2 + 8 \Delta u^2 + 
2 \Gamma^2 \hbar^2 \right) + 
2 t_{\text{SO},1}^2 \left( 2 t_{\text{SO},2}^2 t_{\text{SO},3}^4 + 
t_{\text{SO},2}^4 \left( 16 t_{\text{SO},3}^2 + 
4 \Delta u^2 + \Gamma^2 \hbar^2 \right) \right) \right) \\
&\quad + 32 t_{\text{SO},1}^2 t_{\text{SO},3}^6 \left( 8 t_{\text{SO},1}^4 t_{\text{SO},2}^2 + 
t_{\text{SO},2}^4 \left( 4 \Delta u^2 + \Gamma^2 \hbar^2 \right) + 
t_{\text{SO},1}^2 \left( 8 t_{\text{SO},2}^4 + t_{\text{SO},3}^4 + 
t_{\text{SO},2}^2 \left( 4 (t_{\text{SO},3}^2 + \Delta u^2) + \Gamma^2 \hbar^2 \right) \right) \right) \\
&\quad + 8 t_{\text{C}}^2 t_{\text{SO},3}^4 \left( -48 t_{\text{SO},1}^6 t_{\text{SO},2}^2 + 
3 t_{\text{SO},2}^6 \left( 4 \Delta u^2 + \Gamma^2 \hbar^2 \right) - 
4 t_{\text{SO},1}^2 \left( 5 t_{\text{SO},2}^2 t_{\text{SO},3}^4 + 
t_{\text{SO},2}^4 \left( 24 t_{\text{SO},3}^2 + 
4 \Delta u^2 + \Gamma^2 \hbar^2 \right) \right)\right)\\
&\quad + 8 t_{\text{C}}^2 t_{\text{SO},3}^4 \left(
t_{\text{SO},1}^4 \left( 16 t_{\text{SO},2}^4 + 6 t_{\text{SO},3}^4 - 
t_{\text{SO},2}^2 \left( 32 t_{\text{SO},3}^2 + 12 \Delta u^2 + 
3 \Gamma^2 \hbar^2 \right) \right) \right) \\
& \quad + 4 t_{\text{C}}^4 t_{\text{SO},3}^2 \left( 32 t_{\text{SO},1}^6 t_{\text{SO},2}^2 + 8 t_{\text{SO},2}^8 + 74 t_{\text{SO},2}^4 t_{\text{SO},3}^4 + 
t_{\text{SO},2}^6 \left( 8 t_{\text{SO},3}^2 - 12 \Delta u^2 - 
3 \Gamma^2 \hbar^2 \right)\right)\\
& \quad - 4 t_{\text{C}}^4 t_{\text{SO},3}^2 \left(
4 t_{\text{SO},1}^2 \left( 8 t_{\text{SO},2}^6 + 7 t_{\text{SO},2}^2 t_{\text{SO},3}^4 + 
t_{\text{SO},2}^4 \left( -40 t_{\text{SO},3}^2 + 
4 \Delta u^2 + \Gamma^2 \hbar^2 \right) \right)\right) \\
& \quad + 4 t_{\text{C}}^4 t_{\text{SO},3}^2 \left(
t_{\text{SO},1}^4 \left( -40 t_{\text{SO},2}^4 + 6 t_{\text{SO},3}^4 + 
t_{\text{SO},2}^2 \left( 40 t_{\text{SO},3}^2 + 12 \Delta u^2 + 
3 \Gamma^2 \hbar^2 \right) \right) \right)]
\end{aligned}
\end{equation}
\end{small}

\section{Deriving elliptical quantum dot wave functions}

In this section, we introduce the steps used in deriving the elliptical quantum dot wave functions.
Starting from the left quantum dot (labelled by $a$), which is a circular quantum dot placed in a out-of-plane magnetic field, in the single particle basis, the Hamiltonian read:
\begin{equation}
    H_{a,0} = \frac{(\bm{p}+e\bm{A})^2}{2m_{\text{HP}}} + \frac{1}{2}m_{\text{HP}} \omega_{a,0} (x^2+y^2) \,.
\end{equation}
We consider the gauge to be:
\begin{equation}
    \bm{A} = \pqty{-B_z y,\, B_z x,\,0} \,.
\end{equation}
This Hamiltonian is known as the Fock-Darwin Hamiltonian, which is solved by the following wave-functions for the ground state:
\begin{equation}
    \phi_a(x,y) = \frac{1}{R_a\sqrt{\pi}} \exp(-\frac{x^2+y^2}{2R_{a}^2}) \,.
\end{equation}
The effective quantum dot radius, modified by the magnetic field is:
\begin{equation}
    R_a = \sqrt{\frac{\hbar}{m_{\text{HP}}\Omega_a}} \quad \Omega_a = \sqrt{\omega_0^2+\frac{1}{4}\omega_c^2} \quad \omega_c = \frac{e B_{z}}{m_{\text{HP}}}
\end{equation}
Now, we consider an anisotropic confinement, which describes the a elliptical quantum dot (labeled by $b$):
\begin{equation}
    H_{b,0} = \frac{(\bm{p}+e\bm{A})^2}{2m_{\text{HP}}} + \frac{1}{2}m_{\text{HP}} (\omega_{b,x}x^2+\omega_{b,y}y^2) \,.
\end{equation}
This elliptical Fork-Darwin states can be transformed by the following transformation:
\begin{align}
q_1 &= x \cos(\eta) + A_2 p_y \sin(\eta), \\
q_2 &= y \cos(\eta) + A_2 p_x \sin(\eta), \\
p_1 &= p_x \cos(\eta) - A_1 y \sin(\eta), \\
p_2 &= p_y \cos(\eta) - A_1 x \sin(\eta) \,.
\end{align}
By requiring the commutation relations $\bqty{q_i,\,p_j}=i\hbar\Delta u_{i,j}$, we have $A_1A_2=1$.
We can put the transformed variables in the original Hamiltonian.
The new Hamiltonian has the form:
\begin{equation}
    \frac{p_1^2}{4 m_{\text{HP}}} D_1 + \frac{p_2^2}{4 m_{\text{HP}}} D_2 + \frac{q_1^2}{4 m_{\text{HP}}} D_3 + \frac{q_2^2}{4 m_{\text{HP}}} D_4 + \frac{p_1 q_2}{2 m_{\text{HP}}} D_5 + \frac{p_2 q_1}{2 m_{\text{HP}}} D_6 \,.
\end{equation}
We want the crossing term to vanish ($D_5$=$D_6$=0), which solves the expressions for $A_1$ and $\eta$:
\begin{equation}
    A_1 = \frac{m_{\text{HP}} \sqrt{\Omega_{b,x}^2 + \Omega_{b,y}^2}}{\sqrt{2}} \quad \tan(\eta)=\frac{\sqrt{2} \sqrt{\Omega_{b,x}^2 + \Omega_{b,y}^2} \, \omega_c}{\Omega_{b,x}^2 - \Omega_{b,y}^2} \,,
\end{equation}
where
\begin{equation}
    \Omega_{b,x} = \sqrt{\omega_{b,x}^2+\frac{1}{4}\omega_c^2} \quad \Omega_{b,y} = \sqrt{\omega_{b,y}^2+\frac{1}{4}\omega_c^2} \,.
\end{equation}
Now, the coefficients $D_1, D_2, D_3, D_4$ can be expressed in known quantities like $\Omega_{b,x}, \Omega_{b,y}$:
\begin{align}
D_1&=\frac{\Omega_b^2 + \Omega_{bx}^2 + 3 \Omega_{by}^2}{\Omega_{bx}^2 + \Omega_{by}^2}, \\
D_2&=\frac{-\Omega_b^2 + 3 \Omega_{bx}^2 + \Omega_{by}^2}{\Omega_{bx}^2 + \Omega_{by}^2}, \\
D_3&=\frac{1}{2} m_{\text{HP}}^2 \left( \Omega_b^2 + 3 \Omega_{bx}^2 + \Omega_{by}^2 \right), \\
D_4&=\frac{1}{2} m_{\text{HP}}^2 \left( -\Omega_b^2 + \Omega_{bx}^2 + 3 \Omega_{by}^2 \right) \,,
\end{align}
where
\begin{equation}
   \Omega_b^2 = \sqrt{(\Omega_{bx}^2 - \Omega_{by}^2)^2 + 2 (\Omega_{bx}^2 + \Omega_{by}^2) \omega_c^2}
\end{equation}
The Hamiltonian in new variables are:
\begin{equation}
    H_{b,0} = D_1 \frac{p_1^2}{m_{\text{HP}}} + D_2 \frac{p_2^2}{m_{\text{HP}}} + D_3 \frac{q_1^2}{4m_{\text{HP}}} + D_4 \frac{q_2^2}{4m_{\text{HP}}}
\end{equation}
This Hamiltonian has the solution:
\begin{equation}
    \phi_b(x,y) = \frac{1}{\sqrt{\pi}\sqrt{R_{b,x}}\sqrt{R_{b,y}}} \exp(-\frac{x^2}{2R^2_{b,x}}-\frac{y^2}{2R^2_{b,y}}-i\frac{xy}{R^2_{b,xy}}) \,,
\end{equation}
where:
\begin{equation}
    R_{b,x} = \sqrt{\frac{\hbar D_5}{D_2 D_3}} \quad R_{b,y} = \sqrt{\frac{\hbar D_5}{D_2 D_3}} \quad R_{b,xy}=\sqrt{\frac{\hbar D_5}{(A_1D_1D_2-A_2D_3D_4)\sin\eta}} \,,
\end{equation}
and
\begin{equation}
    D_5 = D_1 D_2 \cos^2\eta + A_2^2 D_3 D_4 \sin^2\eta
\end{equation}
\twocolumngrid
\newpage
%

\end{document}